\RequirePackage{lineno}
\documentclass[twocolumn,letterpaper,aps,prc,longbibliography,superscriptaddress,nofootinbib,floatfix]{revtex4-1}

\usepackage{amsmath}	
\usepackage{graphicx}	
\usepackage{hyperref}
\hypersetup{colorlinks=true, urlcolor=cyan, linkcolor=blue, citecolor=blue}

\usepackage{xspace}	

\newcommand{\jpsi}{\mbox{$J/\psi$}\xspace}

\newcommand{\trento}{\mbox{$\mathrm{T_{R}ENTo}$}\xspace}
\newcommand{\sigabs}{\mbox{$\sigma_{\mathrm{abs}}$}\xspace}
\newcommand{\tform}{\mbox{$\tau_\mathrm{form}$}\xspace}
\newcommand{\sqsn}{\mbox{$\sqrt{s_{_\mathrm{NN}}}$}\xspace}
\def\Ups#1{\mbox{$\Upsilon(#1 S)$}\xspace}
\def\pt{\mbox{$p_T$}\xspace}

\begin{document}

\title{Investigation of suppression of $\Upsilon(nS)$ in relativistic heavy-ion collisions \\at RHIC and LHC energies}
%

\newcommand{\cern}{European Organization for Nuclear Research (CERN), Geneva 01631, Switzerland}
\affiliation{\cern}

\newcommand{\colorado}{Physics Department, University of Colorado, Boulder, Colorado 80309, USA}
\affiliation{\colorado}

\newcommand{\korea}{Department of Physics, Korea University, Seoul 02841, South Korea}
\affiliation{\korea}

\newcommand{\yonsei}{Department of Physics and Institute of Physics and Applied Physics, Yonsei University, Seoul 03722, Korea}
\affiliation{\yonsei}

\newcommand{\inha}{Department of Physics, Inha University, Incheon 22212, South Korea}
\affiliation{\inha}

\newcommand{\jeonbuk}{Division of Science Education, Jeonbuk National University, Jeonju 54896, South Korea}
\affiliation{\jeonbuk}

\newcommand{\pusan}{Department of Physics, Pusan National University, Busan 46241, South Korea}
\affiliation{\pusan}

\newcommand{\sejong}{Department of Physics, Sejong University, Seoul 05006, South Korea}
\affiliation{\sejong}

\newcommand{\heidelberg}{Institute of Physics, Heidelberg University, Heidelberg 69117, Germany}
\affiliation{\heidelberg}

\author{Junlee Kim} \email{junlee.kim@cern.ch} \affiliation{\cern}
\author{Jaebeom Park} \affiliation{\colorado}
\author{Byungsik Hong} \affiliation{\korea}
\author{Juhee Hong} \affiliation{\yonsei}
\author{Eun-Joo Kim} \affiliation{\jeonbuk}
\author{Yongsun Kim} \affiliation{\sejong}
\author{MinJung Kweon} \affiliation{\inha}
\author{Su Houng Lee} \affiliation{\yonsei}
\author{Sanghoon Lim} \email{shlim@pusan.ac.kr} \affiliation{\pusan} 
\author{Jinjoo Seo} \affiliation{\heidelberg}

\date{\today}

\begin{abstract}

The primary purpose of studying quarkonium production in relativistic heavy-ion collisions is to understand the properties of the quark-gluon plasma. At various collision systems, measurements of quarkonium states of different binding energies, such as $\Upsilon(nS)$, can provide comprehensive information. A model study has been performed to investigate the modification of $\Upsilon(nS)$ production in Pb--Pb collisions at $\sqrt{s_{\mathrm{NN}}}=$~5.02~TeV and Au--Au collisions at $\sqrt{s_{\mathrm{NN}}}=$~200~GeV. The Monte-Carlo simulation study is performed with a publicly available hydrodynamic simulation package for the quark-gluon plasma medium and a theoretical calculation of temperature-dependent thermal width of $\Upsilon(nS)$ considering the gluo-dissociation and inelastic parton scattering for dissociation inside the medium. In addition, we perform a systematic study with different descriptions of initial collision geometry and formation time of $\Upsilon(nS)$ to investigate their impacts on yield modification. The model calculation with a varied parameter set can describe the experimental data of $\Upsilon(nS)$ in Pb--Pb collisions at 5.02 TeV and $\Upsilon(2S)$ in Au--Au collisions at 200 GeV but underestimates the modification of $\Upsilon(1S)$ at the lower collision energy. The nuclear absorption mechanism is explored to understand the discrepancy between the data and simulation.


\end{abstract}


\maketitle

\section{Introduction}
\label{sec:Intro}

Quarkonium, which is composed of a heavy quark and its anti-quark, is one of powerful probes to study deconfined and strongly interacting matter, which is so-called quark-gluon plasma (QGP), under the condition of high temperature and high energy density~\cite{Busza:2018rrf, Karsch:2000ps, Shuryak:1977ut, Matsui:1986dk, Digal:2001ue}. During the initial stages of collisions, quarkonium states are produced through hard parton scatterings. These states then undergo the full space-time evolution of the medium. The spectral functions of these states are modified due to a color screening effect~\cite{Matsui:1986dk,Digal:2001ue}, and they also interact with medium constituents such as gluo-dissociation or Landau damping~\cite{laine:2007,Brambilla:2008cx,Brambilla:2010vq}. As a result of interactions between quarkonia and medium~\cite{Gorenstein:2000ck, Andronic:2007bi, Ravagli:2007xx, blaizot:2016jp}, quarkonium yields are expected to be modified. The magnitude of such modification can differ for different quarkonium states, and it also depends on the size and density of the medium. 

One of the most frequently used experimental observables to quantify the modification of quarkonia in heavy-ion collisions is the nuclear modification factor, which is the yield ratio in heavy-ion collisions to that in proton-proton collisions scaled by the average number of binary collisions. Numerous measurements of the nuclear modification factor have been performed at RHIC and LHC energies~\cite{Andronic:2015wma}. Especially, measurements of quarkonium states of the same quark contents and different binding energies, such as \Ups 1, \Ups 2, and \Ups 3, are useful to study the temperature of the medium because initial-state effects are expected to be similar for these states. A clearly ordered suppression of \Ups n states by their binding energies is seen at both RHIC and LHC energies~\cite{STAR:2022rpk, ALICE:2021UpsForward, ATLAS:2022xso, CMS:2018zza, CMS:2023lfu}. Interestingly, the nuclear modification of \Ups 1 in Au--Au collisions at $\sqsn=200$~GeV and Pb--Pb collisions at $\sqsn=5.02$~TeV are close to each other within the uncertainties. In contrast, the modification of \Ups 2 is stronger at 5.02~TeV. The medium size is similar for two collision systems of different collision energies, while the temperature and energy density are higher for the LHC energy. Since a stronger dissociation effect at 5.02~TeV is expected based on these facts, additional effects could cause the different comparison trends for \Ups 1 and \Ups 2 at RHIC and LHC energies. 

To better understand the suppression of bottomonia, there have been many theoretical approaches, such as Landau damping~\cite{Brambilla:2008cx}, Hard-Thermal Loop (HTL)~\cite{Hong:2019ade}, lattice QCD~\cite{Rothkopf:2011db, Lafferty:2019jpr}, and T-matrix method~\cite{Liu:2017qah}, to investigate heavy-quark potential and resulting modifications of heavy quarkonia. Note that two models, based on an open quantum system plus potential Non-Relativistic QCD~\cite{Brambilla:2022ynh} and a transport model~\cite{Du:2017qkv}, expect a stronger suppression at the LHC energy for both \Ups 1 and \Ups 2. Together with many models in the market, the SHINCHON, which stands for Simulation for Heavy IoN Collision with Heavy-quark and ONia, implements a theory baseline in Ref.~\cite{Hong:2019ade} into the hydrodynamically evolving medium to simulate the modification of $\Upsilon(nS)$ at midrapidity more realistically. The inverse reaction of the bottomonium dissociation can enhance the yield of bound states, the so-called regeneration effect, and the effect is not included in the current simulation framework. Note that the regeneration effect is expected not to be significant for bottomonia because of a small cross-section of bottom quark production. In Ref.~\cite{Kim:2022lgu}, we introduced the simulation framework and presented the nuclear modification and elliptic flow results of \Ups n in $p$--Pb, $p$--O and O--O collisions at $\sqsn=8$~TeV. We further extend the framework to study the modification of \Ups n in heavy-ion collisions at different energies.

In the present paper, the nuclear modification factors for \Ups 1, \Ups 2, and \Ups 3 are calculated in Pb--Pb collisions at $\sqrt{s_{\mathrm{NN}}}=$~5.02~TeV and in Au--Au collisions at $\sqrt{s_{\mathrm{NN}}}=$~200~GeV with various simulation configurations, such as initial collision geometry models and the formation time of quarkonia states.
The present paper incorporates the \trento~\cite{Moreland:2014oya} model to construct the initial collision geometry and the SONIC framework~\cite{Romatschke:2015gxa} to simulate the time evolution of the medium. The theoretical baseline of dissociation for \Ups n~\cite{Hong:2019ade} is overlaid onto the evolving medium, where the thermal width for the dissociation is calculated based on HTL perturbation theory.
We present the model calculation results considering the dissociation effect and compare them with experimental results at two energies. Furthermore, additional suppression from the nuclear absorption~\cite{McGlinchey:2012bp} is introduced to understand the similar suppression of \Ups 1 at RHIC and LHC energies.

\section{Simulation framework}  
\label{sec:Simul}

The framework for the present simulation consists of constructing the initial collision geometry, the hydrodynamic evolution for generating medium profiles at each time, and quarkonia responses with the medium. A detailed description of the simulation framework can be found in Ref.~\cite{Kim:2022lgu}. 

\subsection{Initial condition and hydrodynamic evolution}
\label{sec:hydro}

In the previous study~\cite{Kim:2022lgu}, we utilized the Monte Carlo Glauber (MC-Glauber) framework~\cite{Miller:2007ri} for initial collision geometry. Now we use the \trento~\cite{Moreland:2014oya} framework that can simulate different shapes of initial collision geometry with a reduced thickness parameter ($p$) so that we can explore the initial geometry dependence of $\Upsilon$ modification in heavy-ion collisions. The reduced thickness ($f$) of two nucleon thickness functions of $T_{A}$ and $T_{B}$ with $p$ can be expressed as
\begin{eqnarray}
    f = \left( \dfrac{T_{A}^{p} + T_{B}^{p}}{2} \right)^{(1/p)}.
    \label{eq:trentop}
\end{eqnarray}

Two different collision energies are set with the nucleon-nucleon inelastic cross-section of 42.3 and 70.0~mb for $\sqsn=$ 200~GeV and 5.02~TeV, respectively, with the nucleon width of $\sigma=0.4$~fm. The calculated nuclear thickness distribution from wounded nucleons is converted into the temperature with the scale factor, which tunes the final-state multiplicity to be the same with experimental data~\cite{PHOBOS:2010eyu,ALICE:2015juo,ALICE:2018cpu}.

\begin{figure}[!htb]
    \centering
    \includegraphics[width=0.49\linewidth]{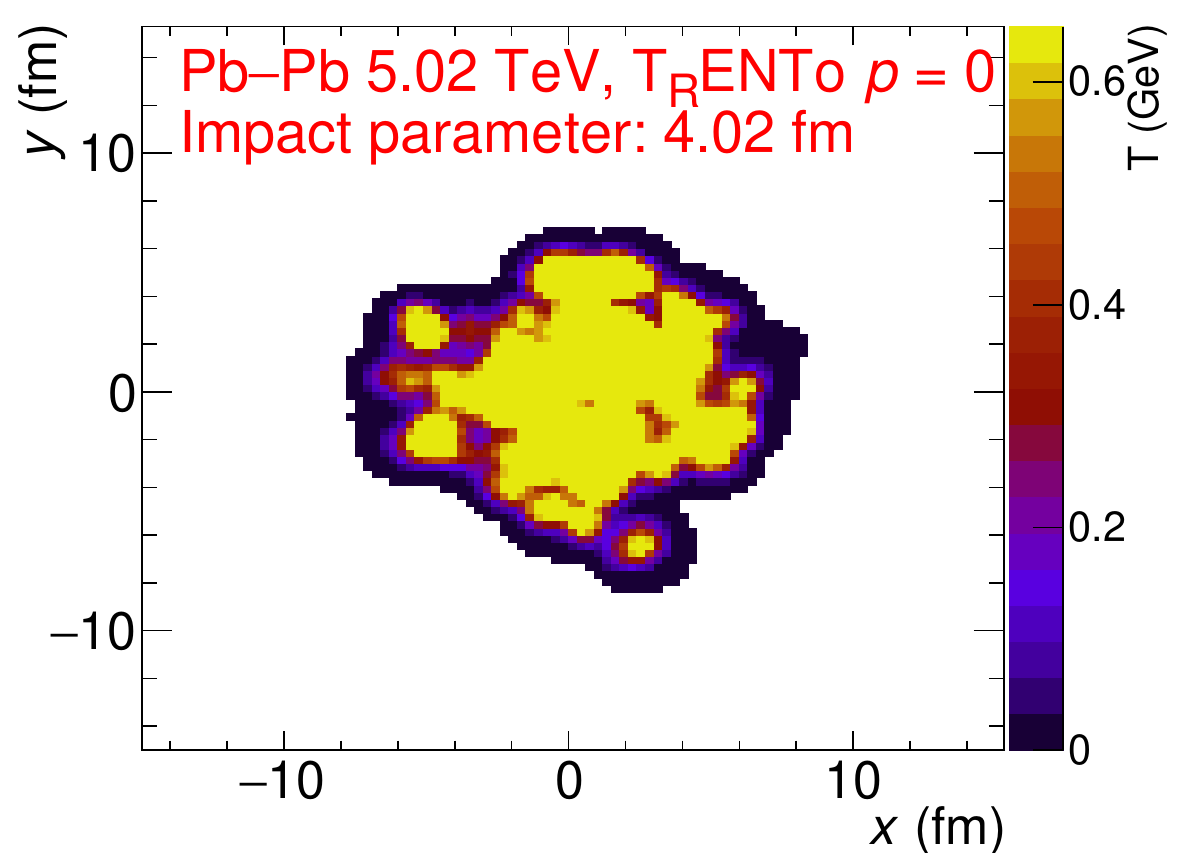}
    \includegraphics[width=0.49\linewidth]{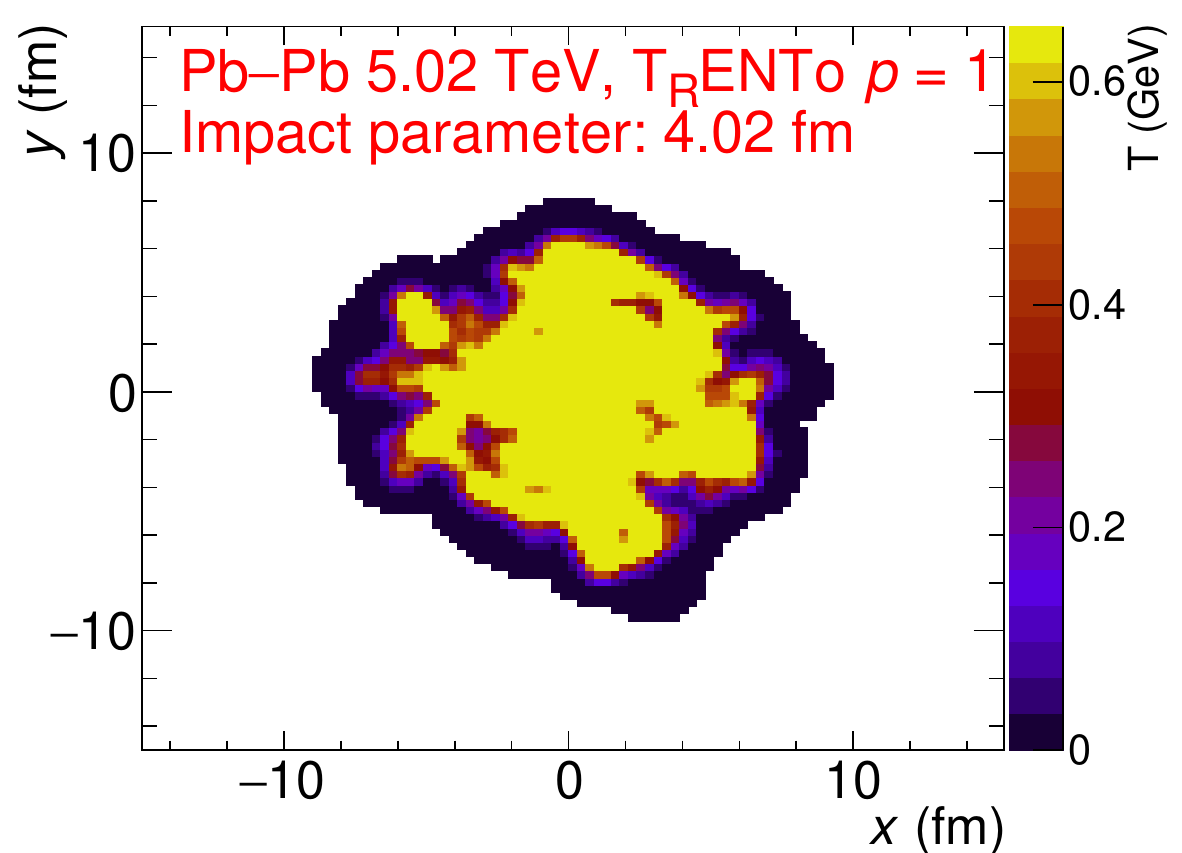}
    \includegraphics[width=0.49\linewidth]{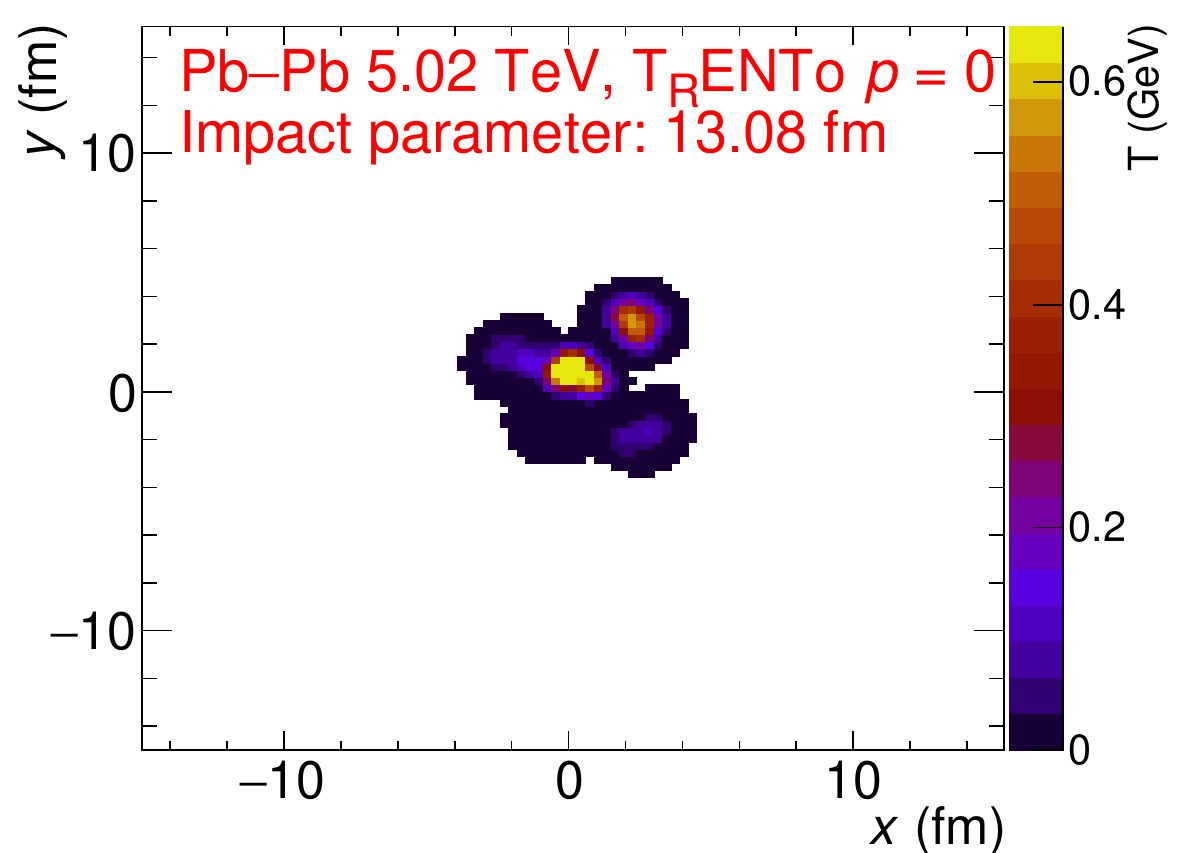}
    \includegraphics[width=0.49\linewidth]{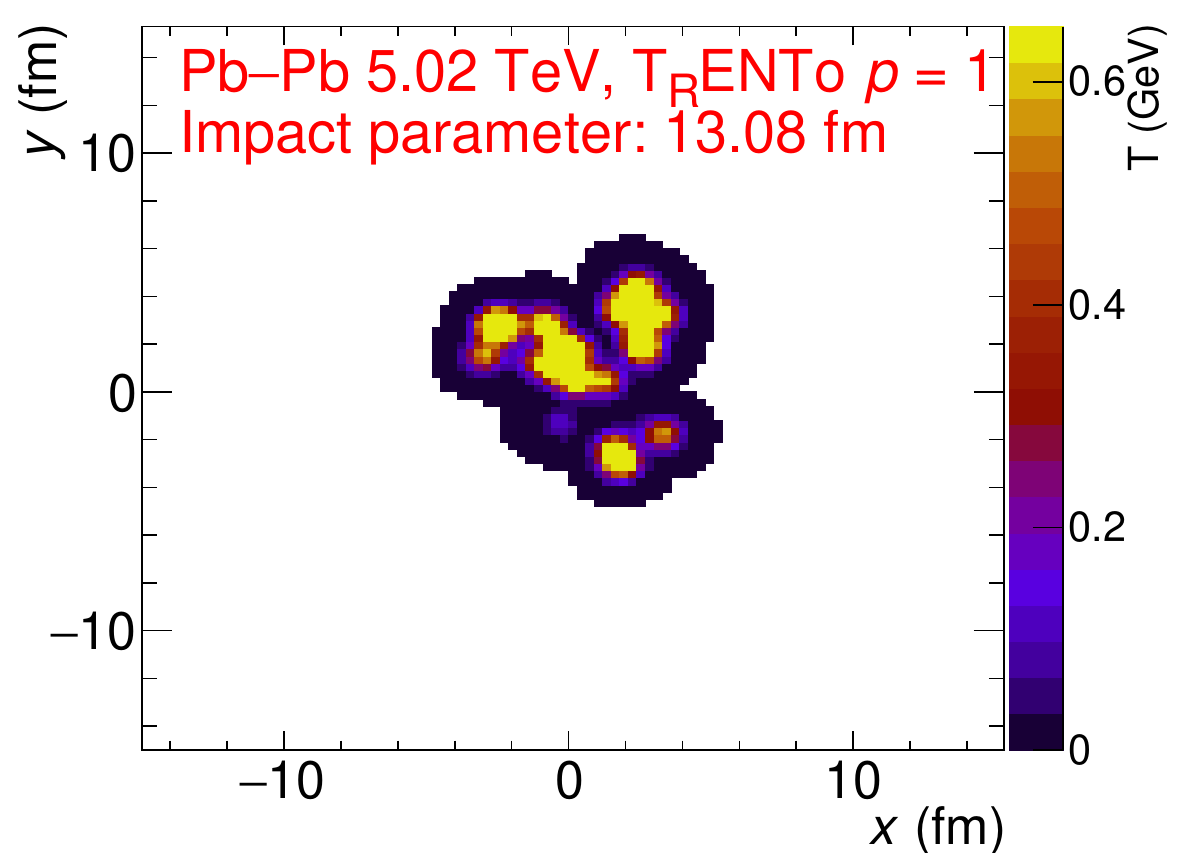}
    
    \includegraphics[width=0.49\linewidth]{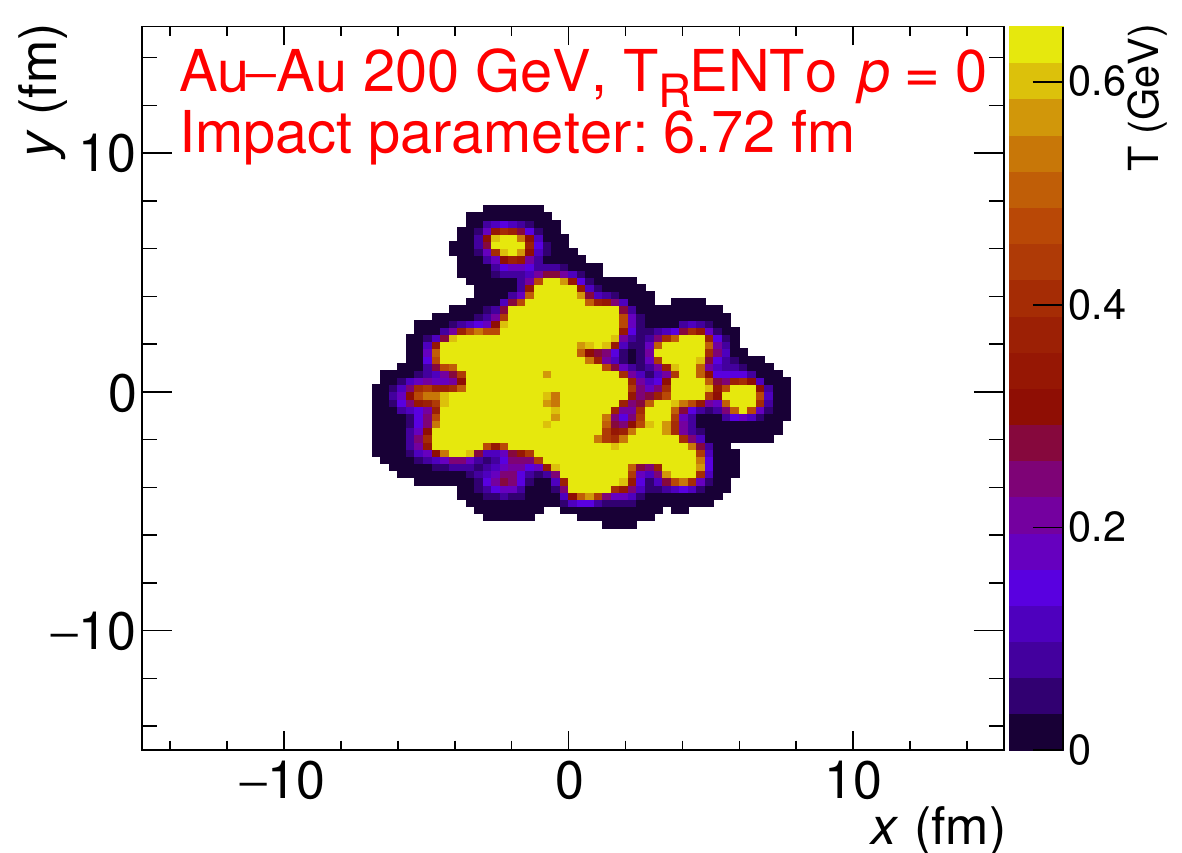}
    \includegraphics[width=0.49\linewidth]{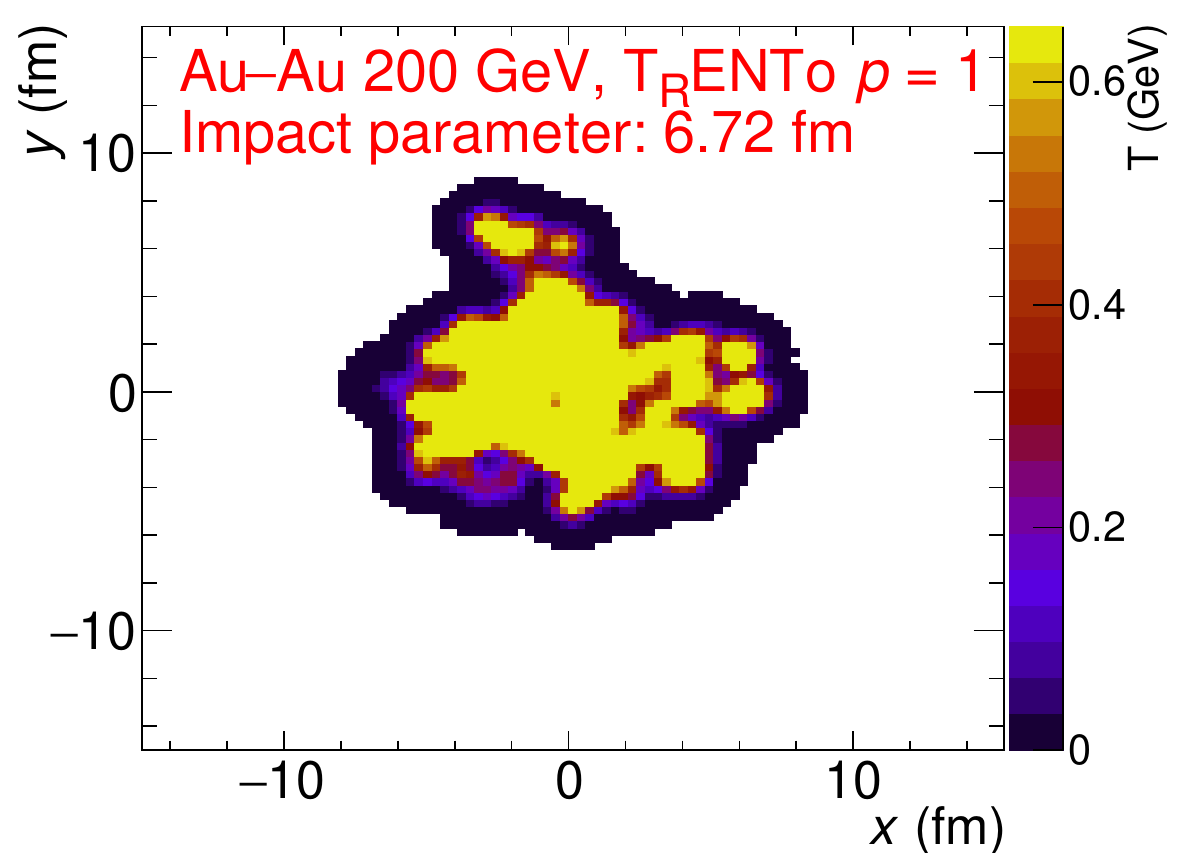}
    \includegraphics[width=0.49\linewidth]{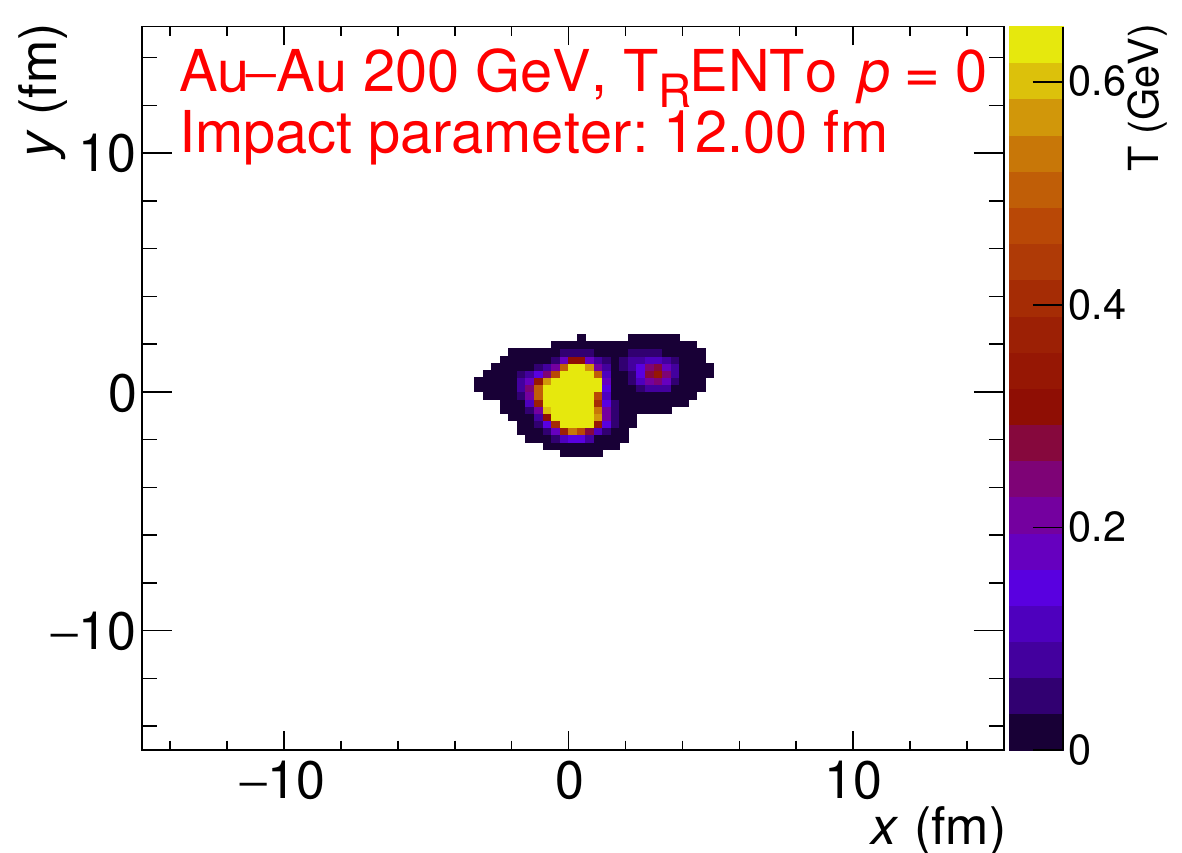}
    \includegraphics[width=0.49\linewidth]{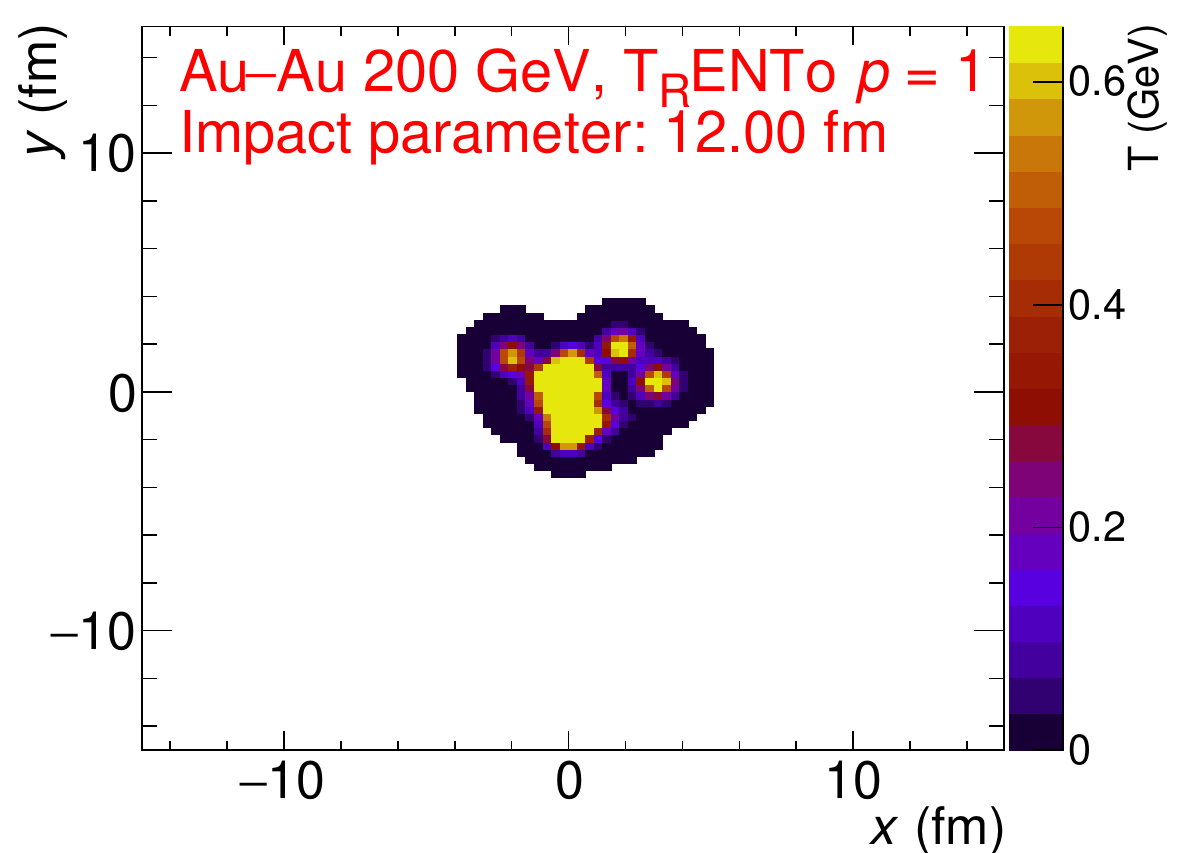}
    \caption{Initial geometries in the temperature unit with different nuclear thickness parameters in the \trento model for central and peripheral heavy-ion collisions.}
    \label{fig:init_shape}
\end{figure}

Figure~\ref{fig:init_shape} shows example events of initial collision geometries in central and peripheral Pb--Pb and Au--Au collisions using \trento with different $p$ parameters. The color indicates the converted temperature for the hydrodynamic simulation. There is no significant difference between $p=0$ and $p=1$ in central collisions, while the thickness parameter affects the initial geometry in peripheral collisions more significantly.

\begin{figure}[!htb]
    \centering
    \includegraphics[width=0.49\linewidth]{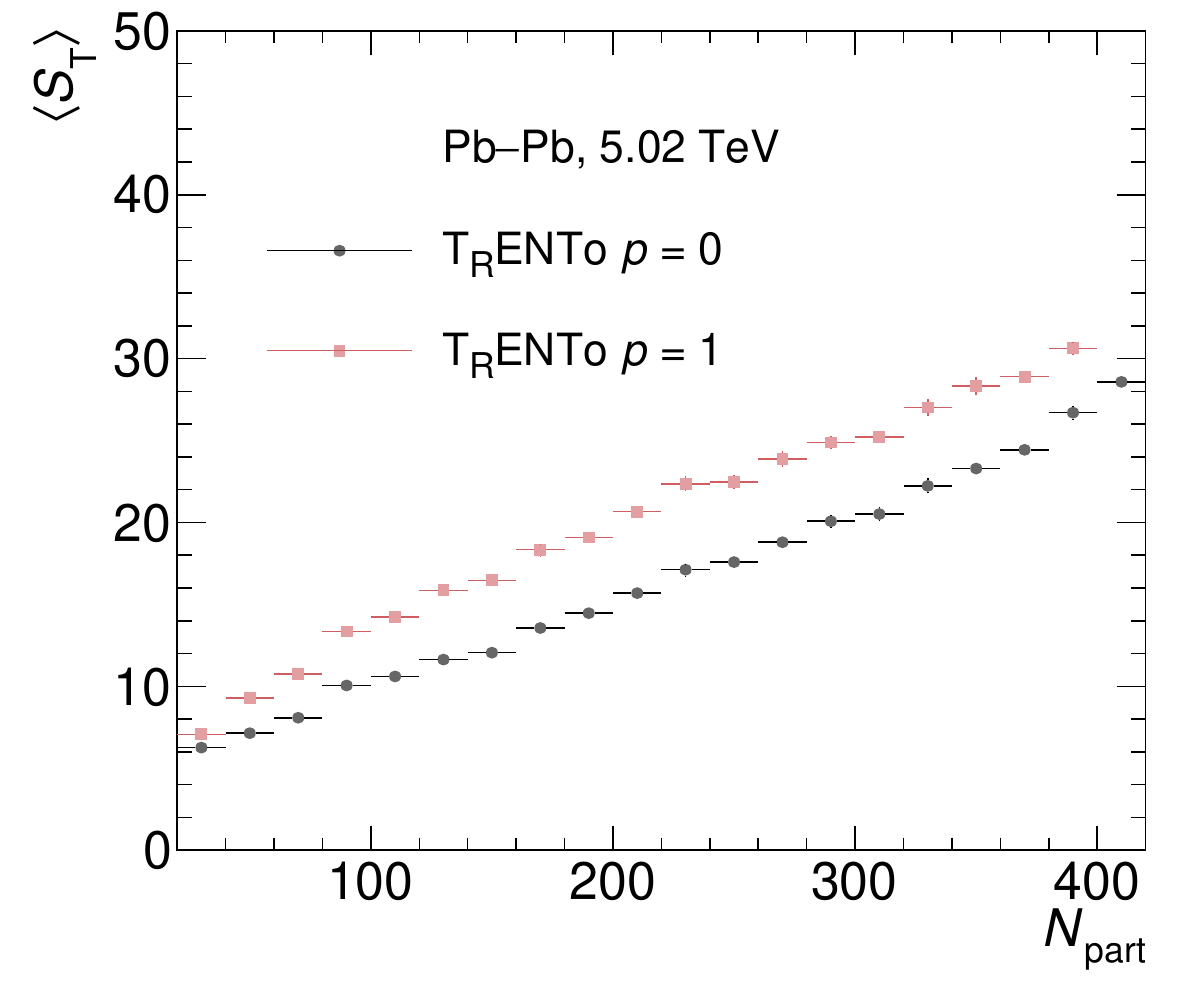}
    \includegraphics[width=0.49\linewidth]{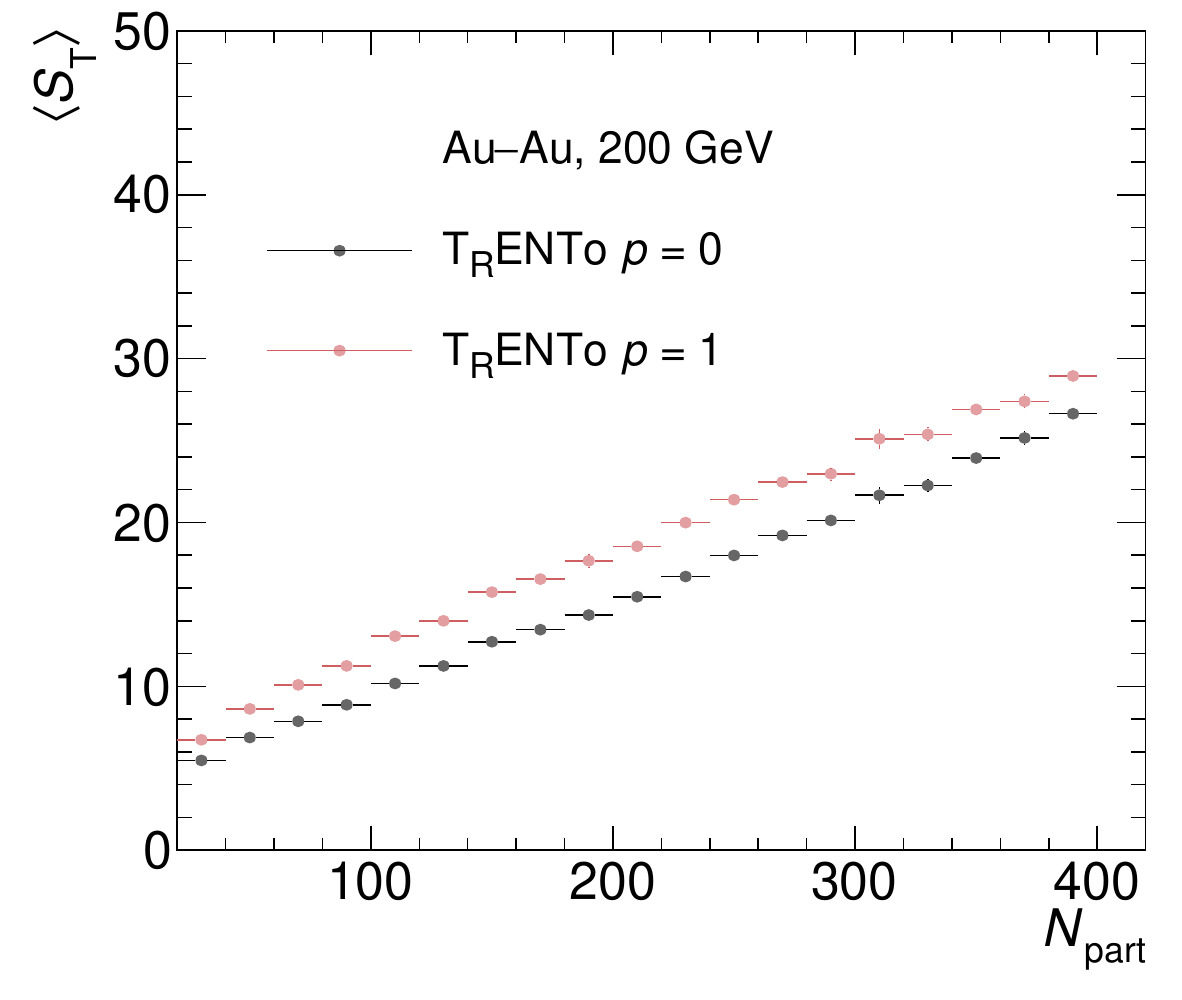}
    \caption{The mean transverse area ($\langle S_T \rangle$) as a function of the number of participants in Pb--Pb collisions and Au--Au collisions for different thickness parameters.}
    \label{fig:fsST}
\end{figure}

For a more quantitative comparison, we calculate the transverse area of the initial geometry defined as 
\begin{eqnarray}
    S_T = \pi \sqrt{\langle x^{2}\rangle \langle y^{2}\rangle - \langle xy\rangle^{2}},
    \label{eq:st}
\end{eqnarray}
where $x$ and $y$ are the spatial coordinates of the participating nucleons in the initial geometry. Figure~\ref{fig:fsST} shows $\langle S_T \rangle$ as a function of the number of participants. A clear difference is seen in mid-central and mid-peripheral collisions. The $\langle S_T \rangle$ with $p=1$, which is the same as the previously used configuration with the MC-Glauber, is larger than the value with $p=0$. The effect of this difference in the area of the medium on the magnitude of \Ups n modification has been studied and will be presented in the following section.

\begin{figure}[!htb]
    \centering
    \includegraphics[width=0.9\linewidth]{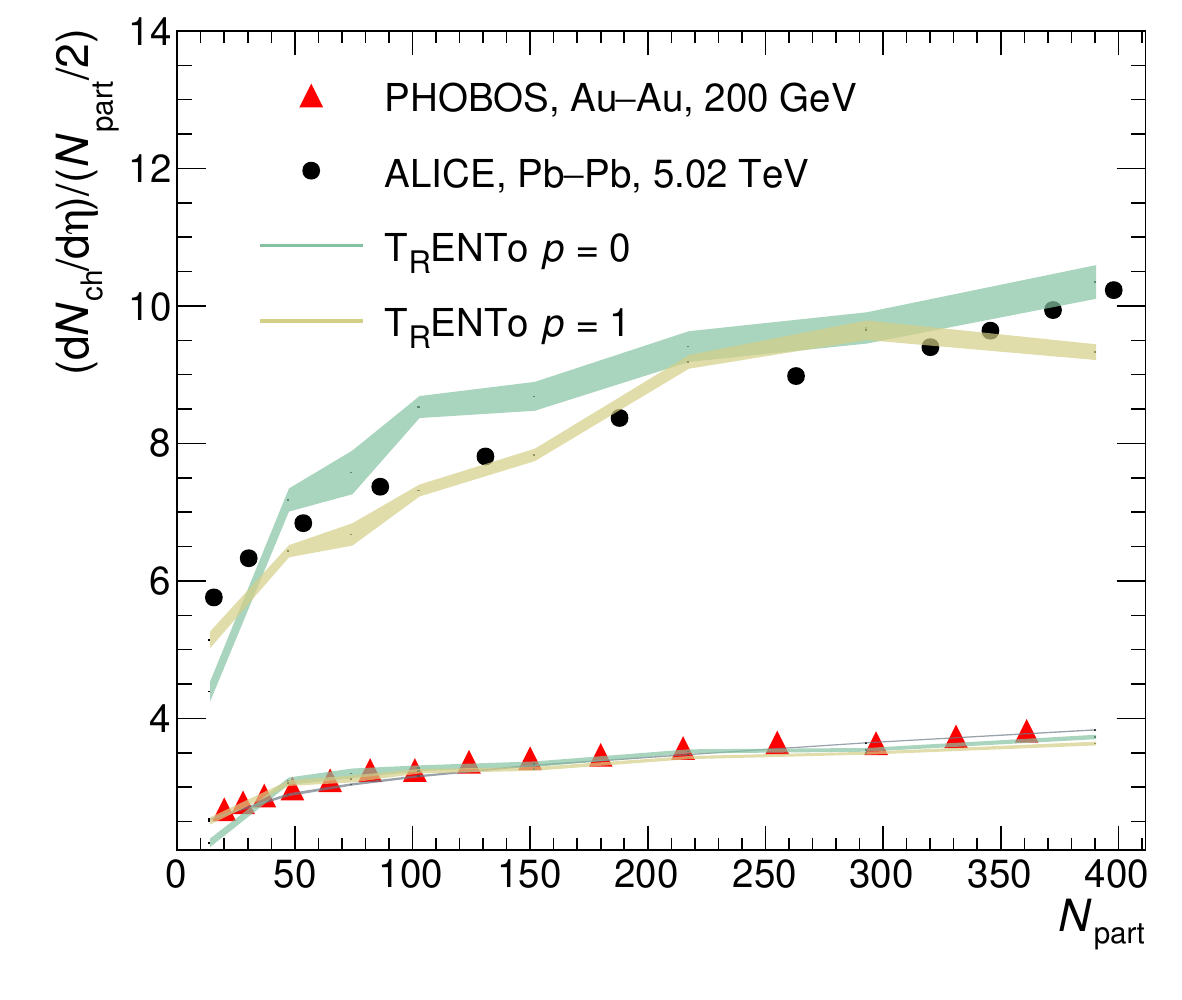}
    \caption{The scaled charged particle multiplicity as a function of the number of participants in Au--Au collisions at $\sqsn=200$~GeV and Pb--Pb collisions at $\sqsn=5.02$~TeV.}
    \label{fig:fsmult}
\end{figure}

Figure~\ref{fig:fsmult} shows the charged particle multiplicity scaled with the number of participants in Au--Au collisions at $\sqsn=$~200~GeV~\cite{PHOBOS:2010eyu} and Pb--Pb collisions at $\sqsn=$~5.02~TeV~\cite{ALICE:2015juo,ALICE:2018cpu}. The hydrodynamic simulation results agree with the experimental measurements throughout the entire centrality range. Note that the scale factors for the initial temperature profile depend on the number of participants for the $p=1$ and are constant for the $p=0$.

\begin{figure}[!htb]
    \centering
    \includegraphics[width=0.49\linewidth]{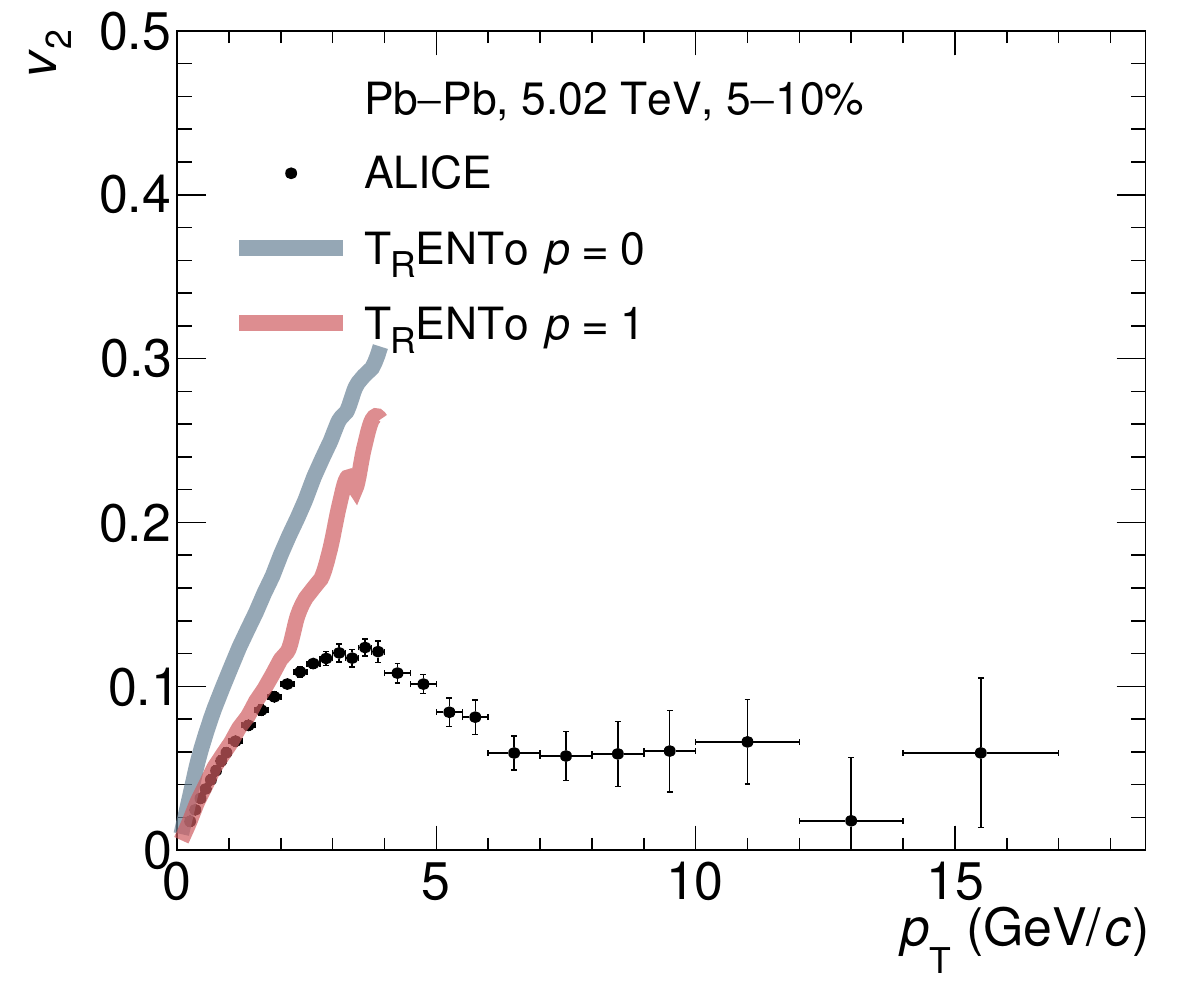}
    \includegraphics[width=0.49\linewidth]{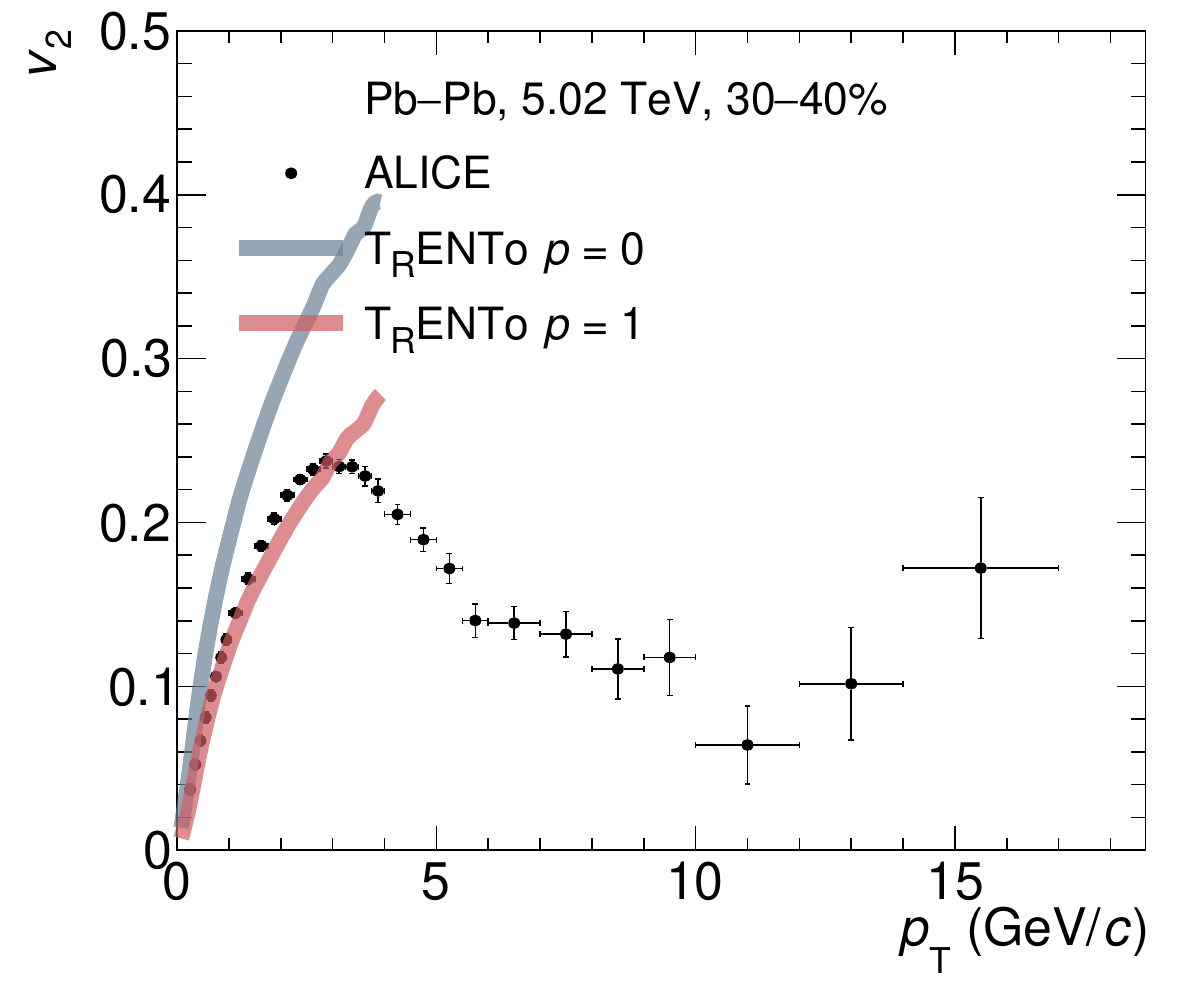}
    \includegraphics[width=0.49\linewidth]{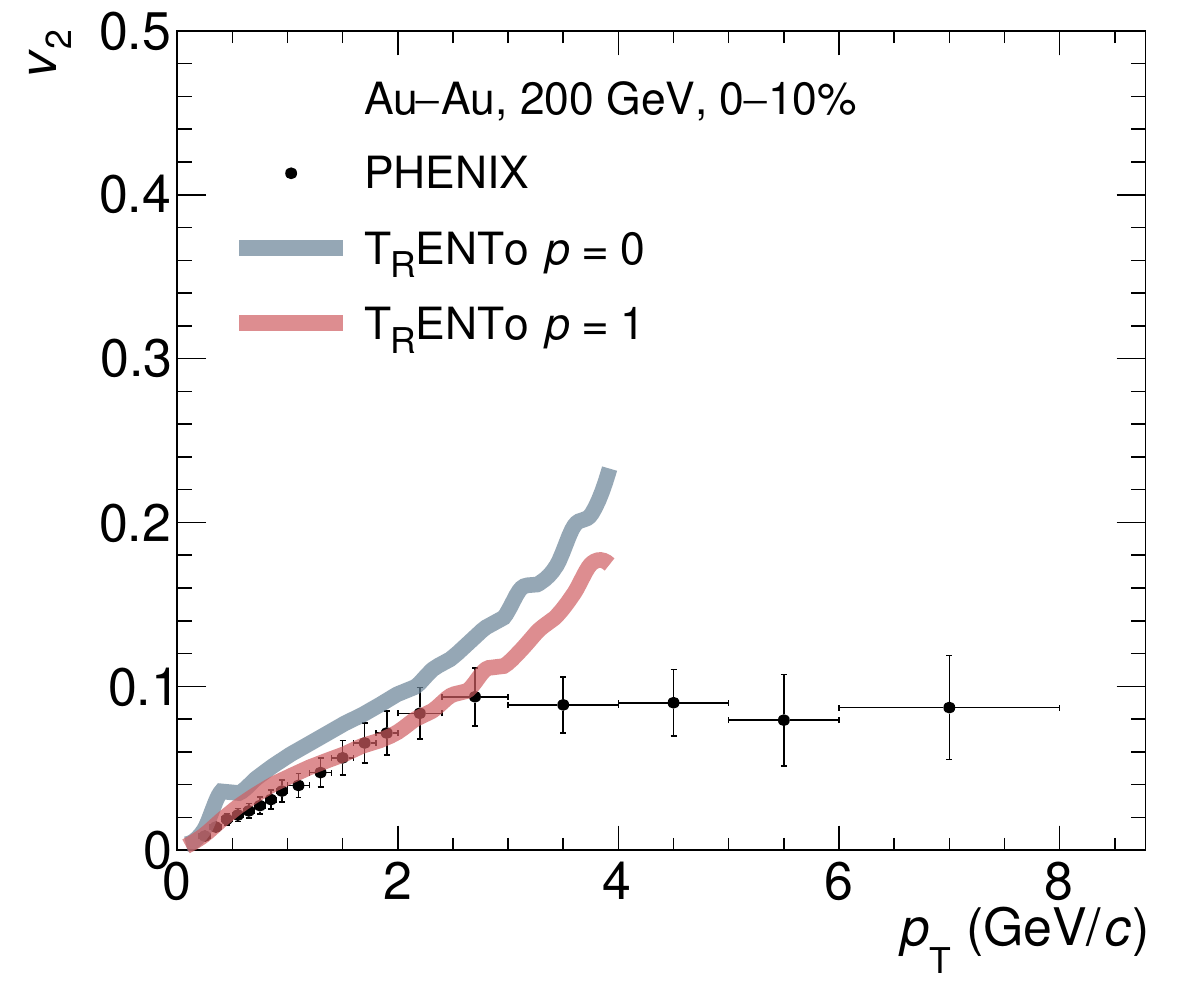}
    \includegraphics[width=0.49\linewidth]{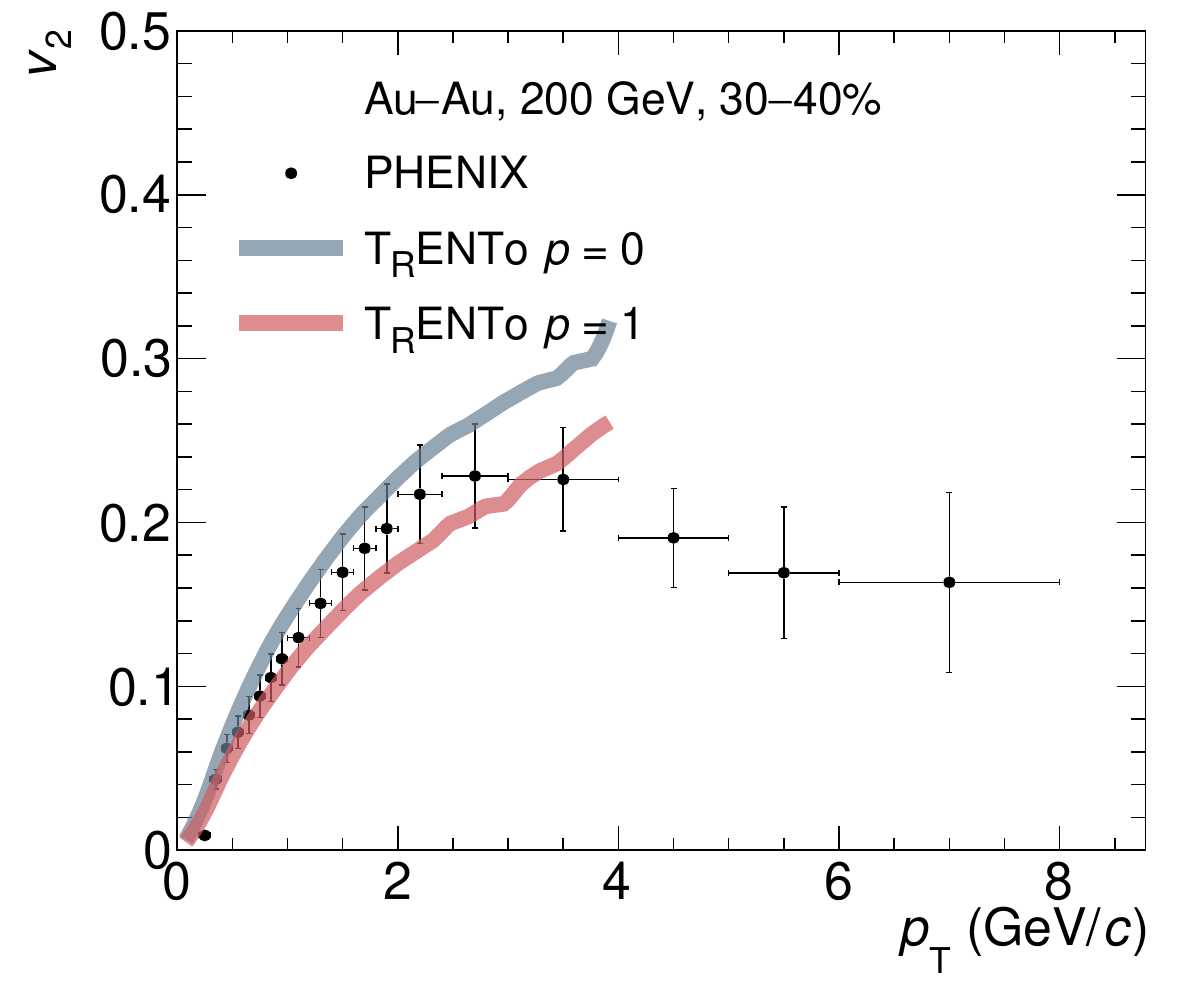}
    \caption{Elliptic flow of charged particles in Pb--Pb collisions at $\sqsn=$~5.02~TeV and Au--Au collisions at $\sqsn=$~200~GeV from the simulation and experimental measurements~\cite{ALICE:2018rtz,PHENIX:2009cjr}. The left (right) panels show the results in central (semi-central) collisions. }
    \label{fig:fsvn}
\end{figure}

For additional check of the background medium and its sensitivity to the $p$ parameter of the \trento model, we also check the elliptic flow coefficients ($v_2$) of charged particles in central and semi-central collisions as shown in Fig.~\ref{fig:fsvn}. The top (bottom) panels represent the results from Pb--Pb (Au--Au) collisions. The simulation results using $p=0$ are slightly larger than those using $p=1$, and both are comparable to the experimental results~\cite{ALICE:2018rtz,PHENIX:2009cjr} at the low transverse momentum ($\pt$) region of $\pt<3~\mathrm{GeV}/c$.


\subsection{Medium responses of $\Upsilon$}
\label{sec:upsres}

The two-dimensional temperature profile of the medium in every time step of the hydrodynamic evolution is obtained from the SONIC simulation. We follow the procedure of medium response described in Ref.~\cite{Kim:2022lgu}. The fraction of survived $\Upsilon$ after a certain time step is calculated as
\begin{eqnarray}
\label{eq:rate}
    \dfrac{N(t+\Delta t,p_{T})}{N(t,p_{T})} = e^{- \int^{t+\Delta t}_{t} d t^{\prime} \Gamma_\mathrm{diss}(t^{\prime},p_{T})},
\end{eqnarray}
where $\Gamma_\mathrm{diss}$ is the thermal width, which depends on the temperature of the medium, the $\Upsilon$'s \pt, and the kind of $\Upsilon$ state, and $\Delta t$ is set to 0.03~fm/$c$ in the hydrodynamic simulation. The thermal width is numerically calculated considering gluo-dissociation and inelastic parton scattering, and more details on the theory can be found in Ref.~\cite{Hong:2019ade}.

The full dissociation temperature, in which the binding energy of the quarkonia reaches zero, is set to 600, 240, and 190~MeV for \Ups 1, \Ups 2, and \Ups 3, respectively, based on Refs.~\cite{MPS:2013qqgp,HS:2006cdqb}. When the medium temperature at $\Upsilon$'s position is higher than the full dissociation temperature, the simulation immediately dissociates the $\Upsilon$. The formation time ($\tform$) of each $\Upsilon$ state is set to 0.5, 1.0, and 1.5~fm/$c$ for \Ups 1, \Ups 2, and \Ups 3, respectively, based on Ref.~\cite{Du:2017qkv}. The actual medium response of each $\Upsilon$ state begins from $t>\gamma \tform$, where $\gamma$ is the Lorentz factor. We have performed a systematic study using various $\tform$ to evaluate how the nuclear modification factor is sensitive to the choice of $\tform$.
Note that there is a recent study to determine the formation time based on various experimental measurements~\cite{Ferreiro:2021kwk}, but it is difficult to draw any firm conclusion due to large experimental uncertainties. 

The feed-down correction is applied to consider the medium response of inclusive bottomonium production by following the procedure described in previous publication~\cite{Kim:2022lgu}. The feed-down correction for the nuclear modification factor can be expressed as
\begin{eqnarray}
    R_{n}(p_T) = \sum_{i} R_{i}(p_T)  \mathcal{F}^{Q_{i}}_{Q_{n}} (p_T),
\end{eqnarray}
where $R_{n}$ is the weighted averaged nuclear modification factor for a certain \Ups n state, $R_{i}$ represents the value for a certain state contributing the the \Ups n state, and $\mathcal{F}^{Q_{i}}_{Q_{n}}$ is the feed-down fraction. The effect of \pt shifts by the decays is not considered due to their similar masses. The nuclear modification factors for $P$-wave states are assumed as $R_{\chi_{b}(1P)} \approx R_{\Upsilon(2S)}$ and $R_{\chi_{b}(2P)} \approx R_{\chi_{b}(3P)} \approx R_{\Upsilon(3S)}$ based on the similarity of binding energies.


\section{Results and Discussions}
\label{sec:Results_Disscussion}

\subsection{Nuclear modification factor}
\label{sec:res_rpa}

\begin{figure}[!htb]
    \centering
    \includegraphics[width=0.99\linewidth]{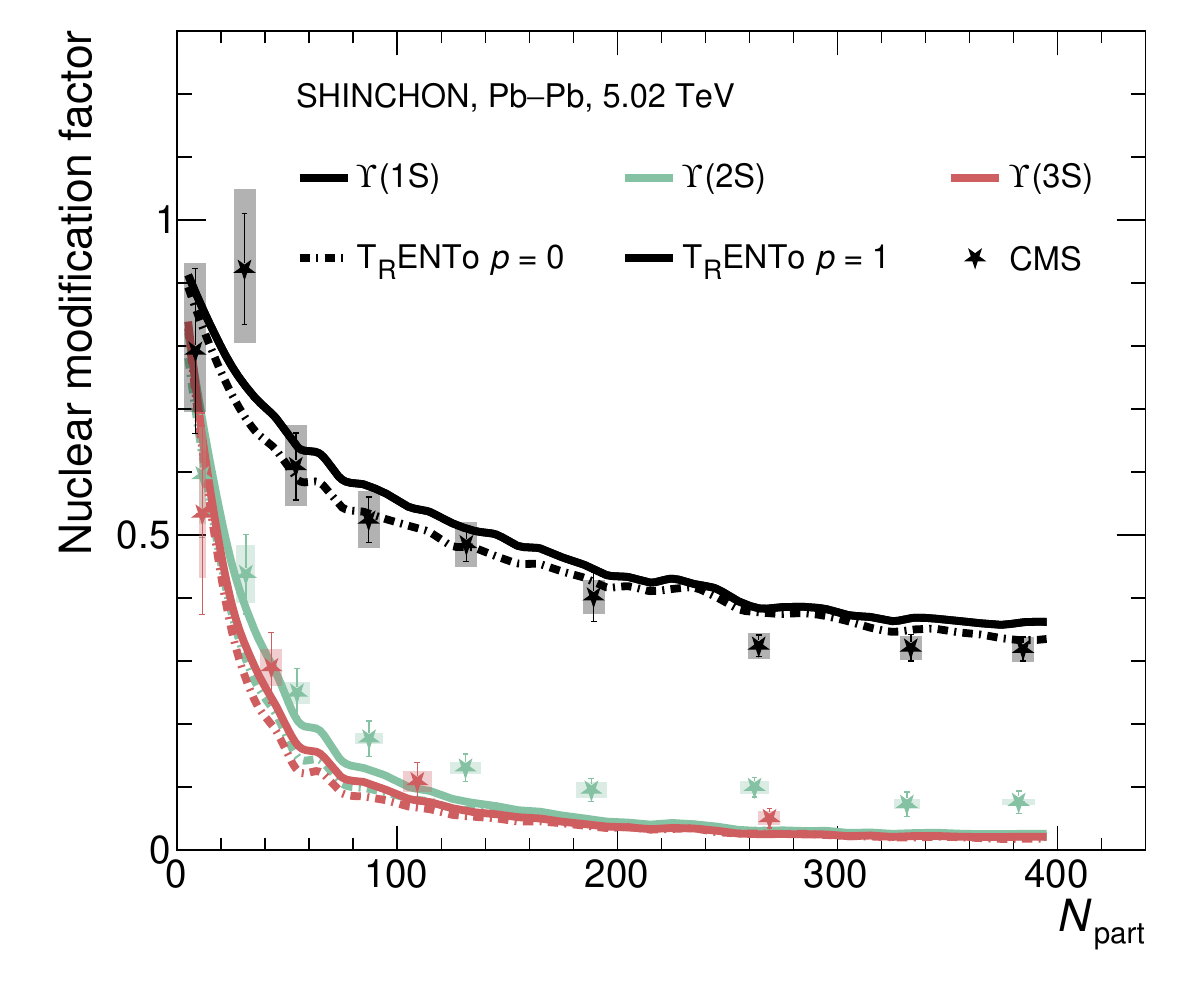}
    \caption{The nuclear modification factors of \Ups n in Pb--Pb collisions at $\sqsn=$~5.02~TeV. Simulation results using two $p$ parameters in the \trento model are compared with the experimental results by CMS~\cite{CMS:2018zza,CMS:2023lfu}. Statistical and systematic uncertainties of the experimental results are represented as vertical lines and boxes, respectively.}
    \label{fig:Comp_NMF_npart_trento_PbPb}
\end{figure}

Figure~\ref{fig:Comp_NMF_npart_trento_PbPb} demonstrates the nuclear modification factor as a function of the number of participants for \Ups 1, \Ups 2, and \Ups 3 mesons in Pb--Pb collisions at $\sqsn=$~5.02~TeV with different $p$ parameter in the \trento model. A slightly stronger modification is seen for the $p=0$ case of a smaller area and higher temperature, but the difference is smaller than the experimental uncertainties. The nuclear modification factor decreases with the increasing number of participants for all states, and an ordered suppression is seen in the model calculation as seen in the experimental results~\cite{CMS:2018zza,CMS:2023lfu}. In addition, the experimental results of \Ups 1 are quantitatively agreed with the model. However, the modifications for \Ups 2 and \Ups 3 are slightly overestimated with the default configuration of the model.

\begin{figure}[!htb]
    \centering
    \includegraphics[width=0.99\linewidth]{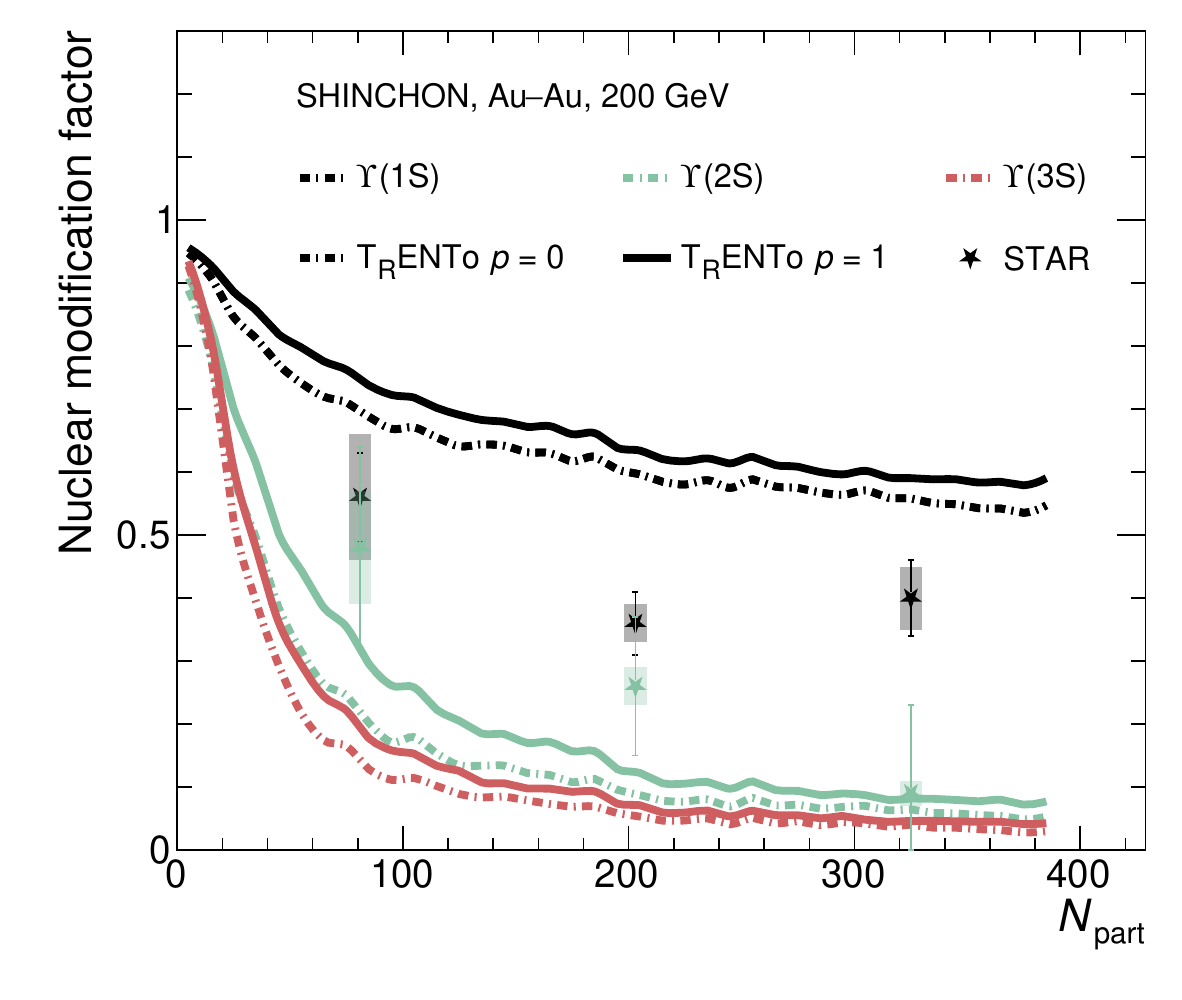}
    \caption{The nuclear modification factors of \Ups n in Au--Au collisions at $\sqsn=$~200~GeV. Simulation results using two $p$ parameters in the \trento model compared with the experimental results by STAR~\cite{STAR:2022rpk}. Statistical and systematic uncertainties of the experimental results are represented as vertical lines and boxes, respectively.}
    \label{fig:Comp_NMF_npart_trento_AuAu}
\end{figure}

\begin{figure}[!htb]
    \centering       
    \includegraphics[width=0.96\linewidth]{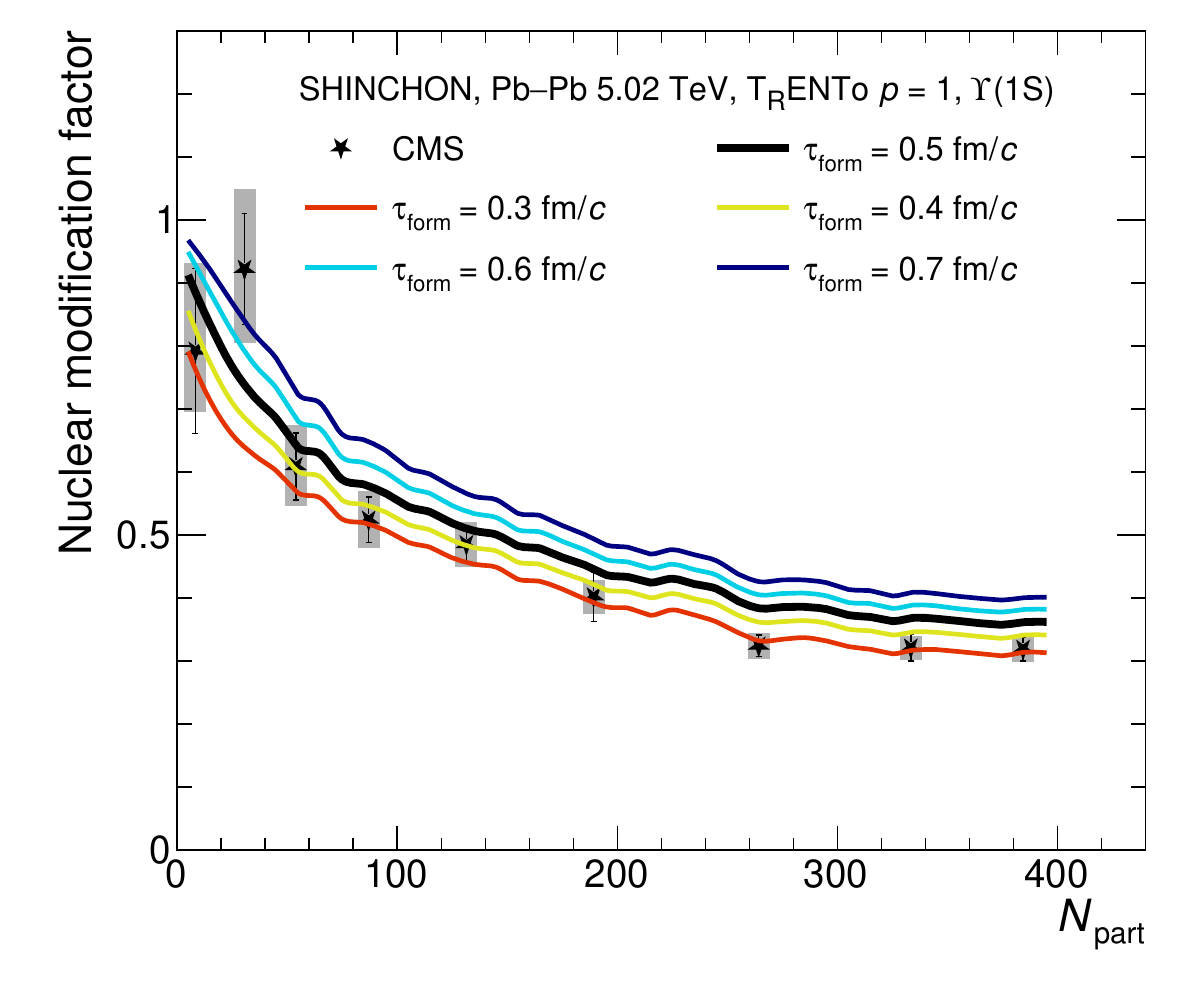}
    \includegraphics[width=0.96\linewidth]{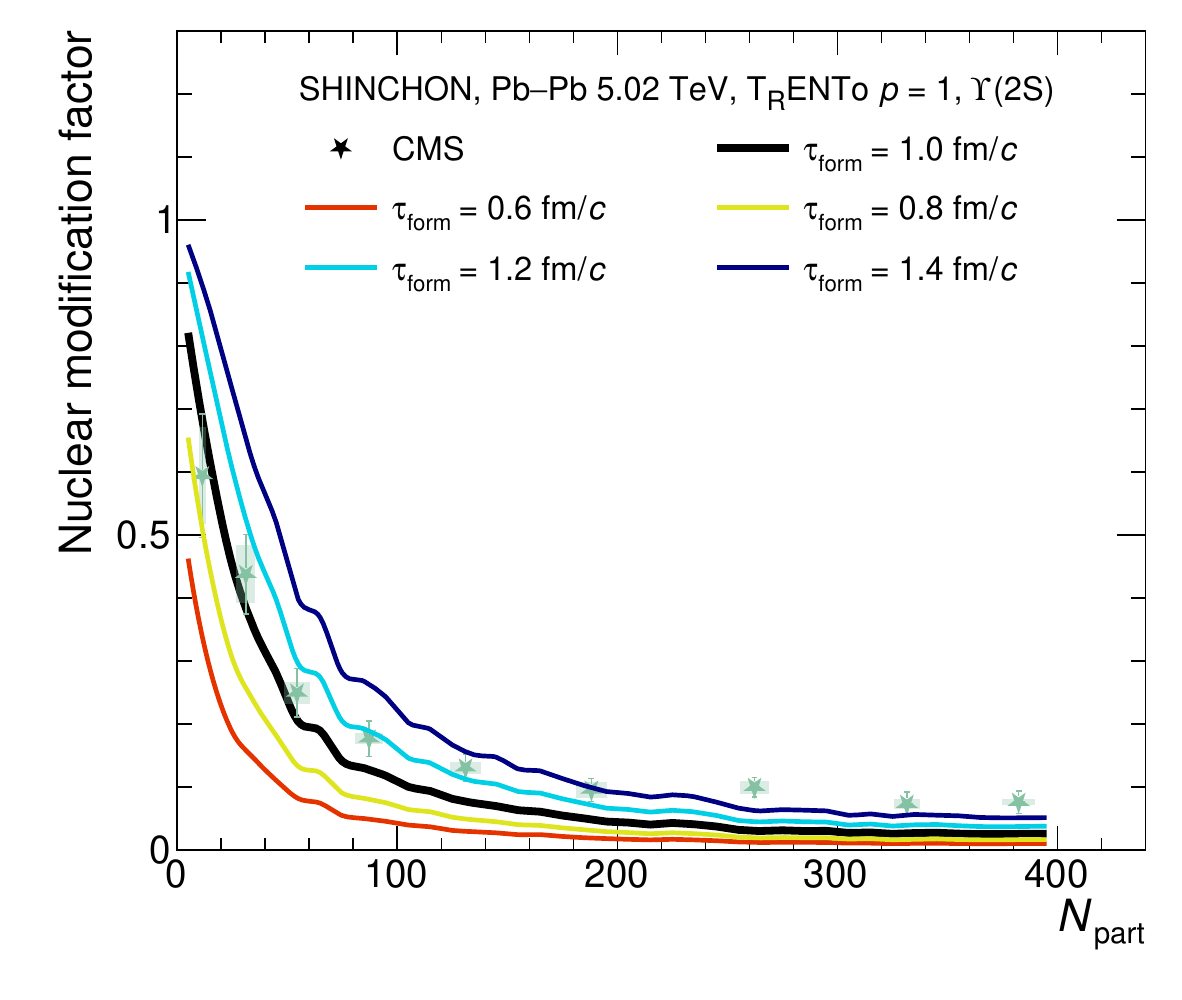}
    \includegraphics[width=0.96\linewidth]{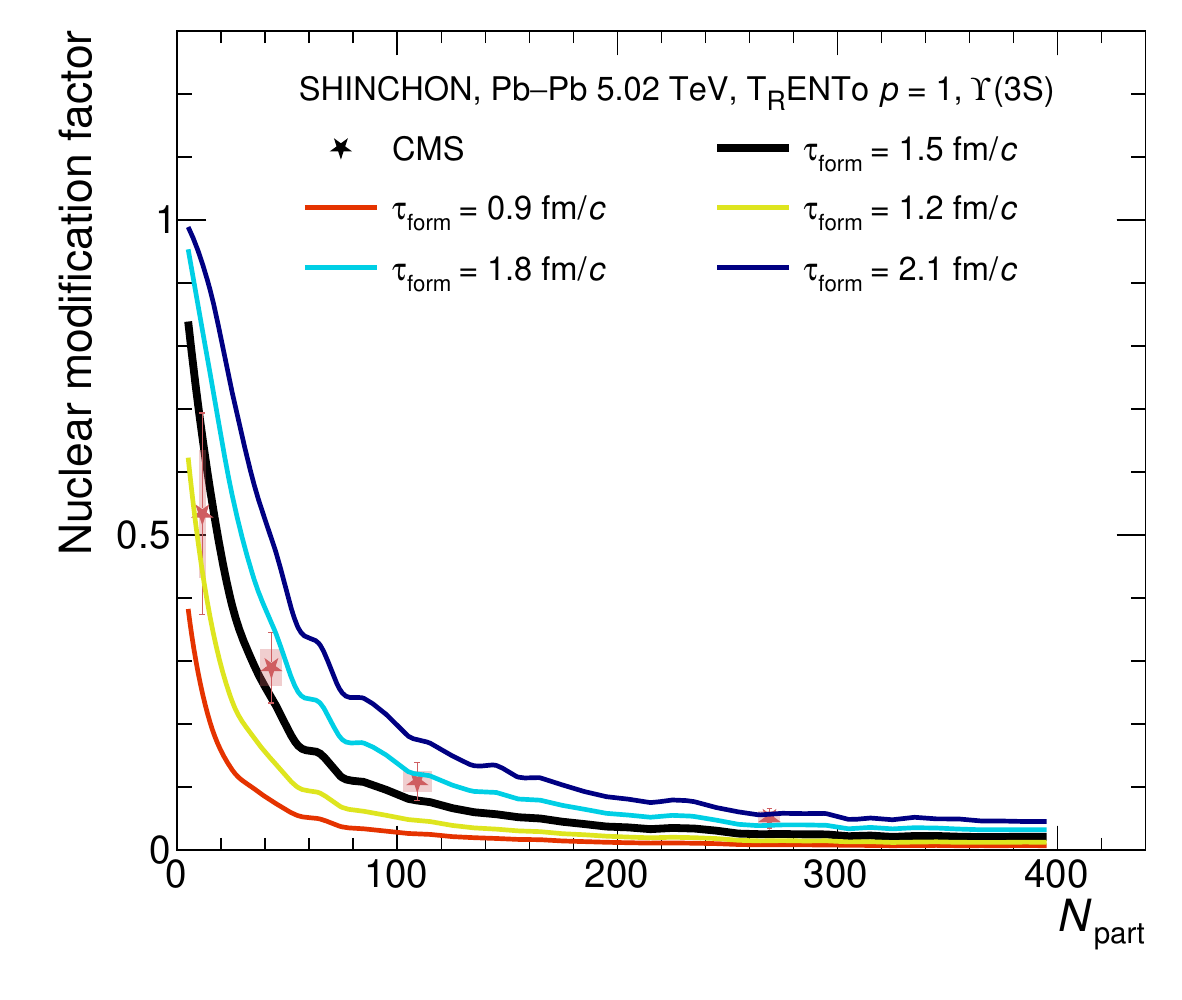}
    \caption{The nuclear modification factor of \Ups n as a function of the number of participants in Pb--Pb collisions at $\sqsn=5.02$~TeV with various formation times.}
    \label{fig:NMF_tform_PbPb}
\end{figure}

\begin{figure}[!htb]
    \centering
    \includegraphics[width=0.96\linewidth]{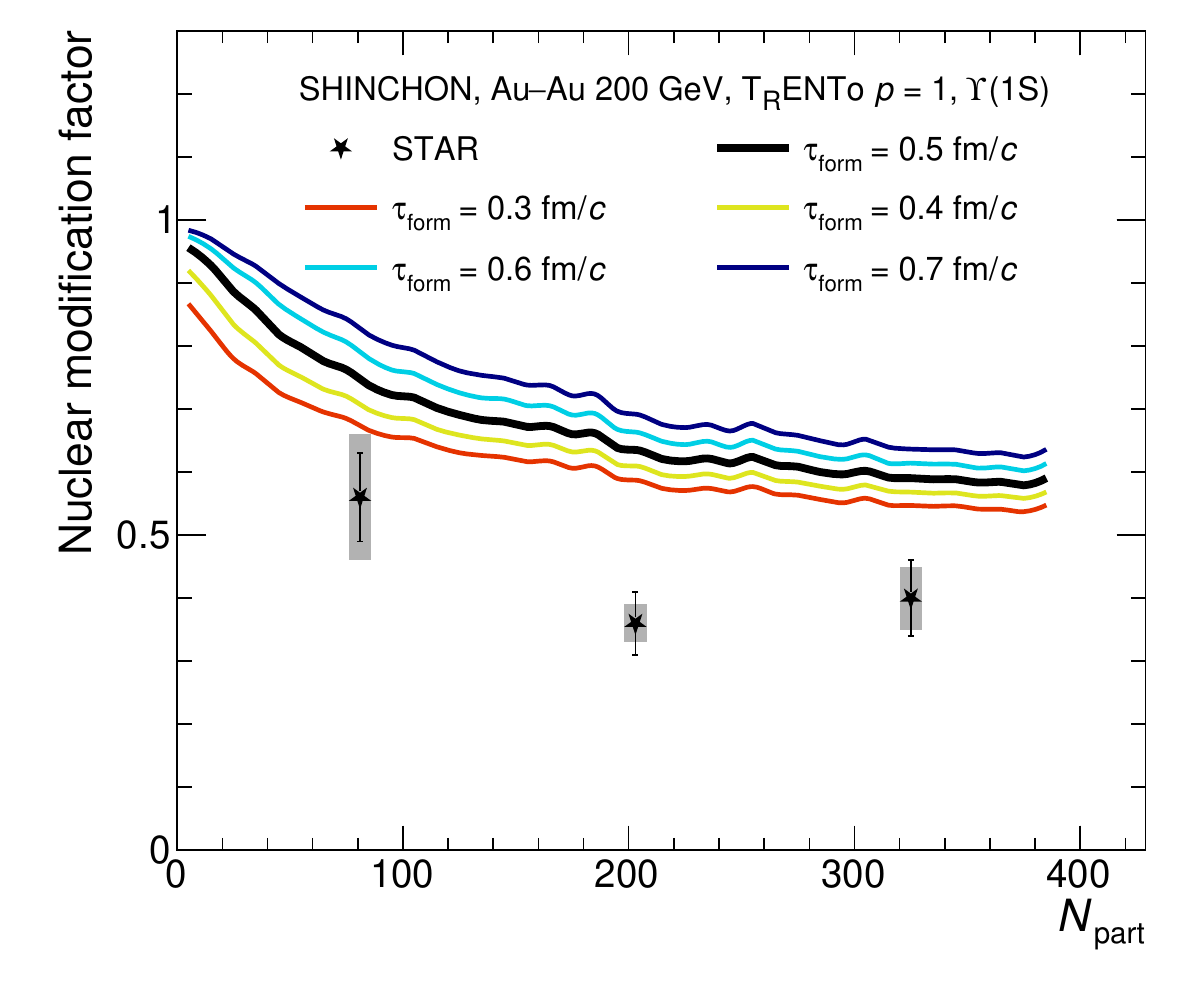}
    \includegraphics[width=0.96\linewidth]{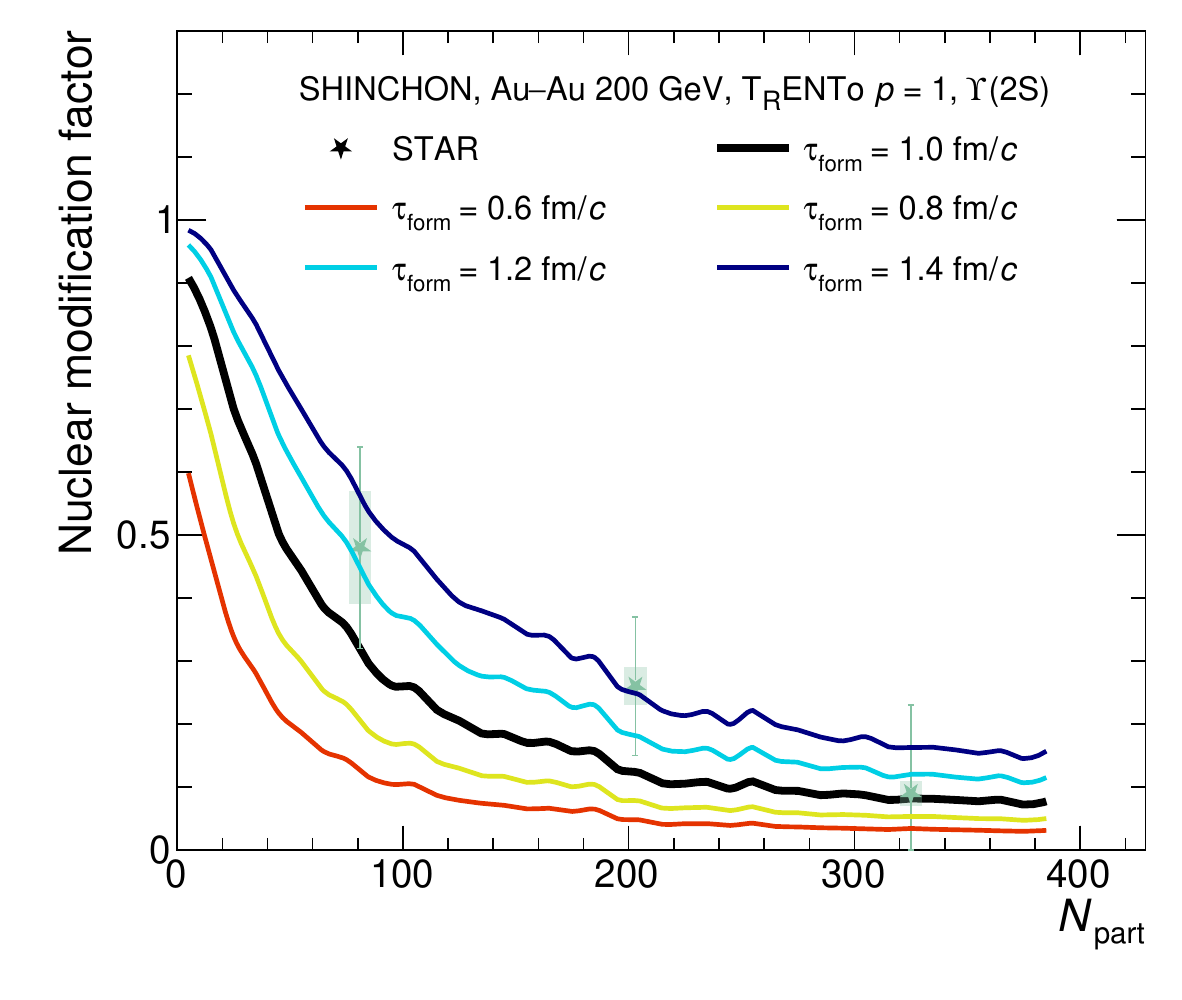}
    \caption{The nuclear modification factor of \Ups n as a function of the number of participants in Au--Au collisions at $\sqsn=200$~GeV with various formation times.}
    \label{fig:NMF_tform_AuAu}
\end{figure}

Figure~\ref{fig:Comp_NMF_npart_trento_AuAu} shows the nuclear modification factors as a function of the number of participants for \Ups 1 and \Ups 2 mesons in Au--Au collisions at $\sqsn=$~200~GeV. The difference between results using two $p$ parameters in the \trento model is larger than those for the higher collision energy but still less significant than experimental uncertainties. Unlike the description for the LHC energy, the model results for the RHIC energy underestimate (overestimate) the modification for \Ups 1 (\Ups 2). When comparing the model results for two energies, the nuclear modification factors of all three states for 5.02~TeV are systematically lower than those for 200~GeV. This trend is consistent with other model calculations presented in Ref.~\cite{STAR:2022rpk}. As pointed out in the introduction, the collision energy dependence seen in various models differs from the trend in the data for \Ups 1 while exhibiting the same trend for \Ups 2.



As a systematic study of the model, we evaluate the effect of the choice of \tform on the nuclear modification factor. In the default configuration, we have used 0.5, 1.0, and 1.5~fm/$c$ for \Ups 1, \Ups 2, and \Ups 3, respectively. We vary the \tform by $\pm0.2$, $\pm0.4$, and $\pm0.6$~fm/$c$ for \Ups 1, \Ups 2, and \Ups 3, respectively. Figure~\ref{fig:NMF_tform_PbPb} shows the nuclear modification factor for \Ups n in Pb--Pb collisions at 5.02~TeV with various \tform parameters. The modification is getting more significant as \tform decreases because the duration of medium response of $\Upsilon$ increases. The model results show a clear dependence on the choice of \tform. The CMS results are covered by the systematic sets of \tform for all three states. The \Ups 1 (\Ups 2 and \Ups 3) results favor the model calculation with a smaller (larger) \tform parameter.

Figure~\ref{fig:NMF_tform_AuAu} shows the nuclear modification factor for \Ups n in Au--Au collisions at 200~GeV with various \tform parameters. In the case of \Ups 1, all nuclear modification factors with various \tform are located above the STAR results, which differs from the comparison at 5.02~TeV. On the other hand, the model calculation with $\tform=1.2$~fm/$c$ for \Ups 2 provides a better description of the experimental data than that with $\tform=1.0$~fm/$c$, which is the same case as the higher energy. This observation may suggest additional suppression mechanisms for \Ups 1 at $\sqsn=$~200~GeV.




\subsection{Nuclear absorption}
\label{sec:res_rpa_abs}
The modification of quarkonia yield at fixed target experiments is explained by a breakup of bound states in collisions with nucleons inside the target~\cite{Arleo:1999af,NA50:2006rdp}. Although the breakup effect becomes smaller in higher collision energy due to a shorter nucleus crossing time compared to the formation time of quarkonia, a significant breakup effect was necessary to describe the suppression of $\jpsi$ and $\Upsilon$ at backward rapidity compared to the modification expected from the nuclear parton distribution in $d$--Au collisions at $\sqsn=200$~GeV~\cite{PHENIX:2010hmo,PHENIX:2012xws}. An extensive model study was done to extract the effective absorption cross-section (\sigabs) based on the nuclear modification factor of \jpsi~\cite{McGlinchey:2012bp}, and the extracted cross-section is about 2~mb at mid-rapidity and 5~mb at backward rapidity. At the LHC energy, where the beam crossing time is much shorter, the nuclear modification factor of \jpsi at backward rapidity can be described by the model considering only nuclear parton distribution~\cite{ALICE:2015kgk}. 


\begin{figure}[!htb]
    \centering       
    \includegraphics[width=0.99\linewidth]{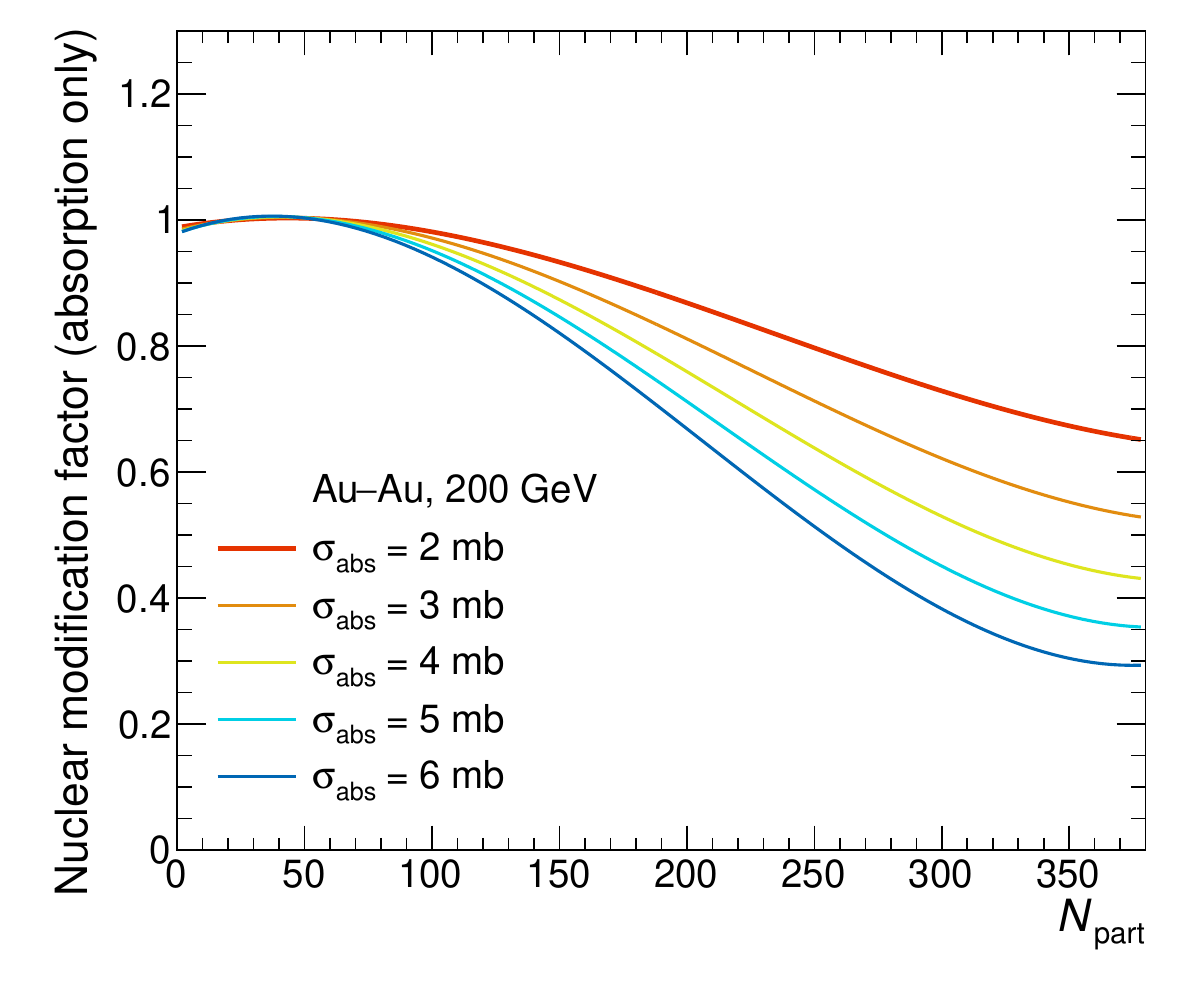}
    \caption{The nuclear modification factor for given nuclear absorption cross-section (\sigabs) as a function of the number of participants in Au--Au collisions.}
    \label{fig:NMF_absonly}
\end{figure}

We incorporate the nuclear breakup effect in the model calculation for the RHIC energy by following the procedure in Ref.~\cite{McGlinchey:2012bp}. Since there is a lack of experimental measurements of $\Upsilon$ in $p/d$--$A$ collisions to extract the \sigabs, we test various \sigabs values as done in Ref.~\cite{PHENIX:2012xws}. Figure~\ref{fig:NMF_absonly} shows the nuclear modification factor solely from the nuclear absorption as a function of the number of participants in Au--Au collisions for different \sigabs from 2~mb to 6~mb. Note that the magnitude of modification for a given \sigabs is twice larger in Au--Au than in $p/d$--Au when considering interactions with nucleons in both nuclei. The suppression is getting monotonously stronger with increasing \sigabs and the number of participants.

\begin{figure}[!htb]
    \centering       
    \includegraphics[width=0.99\linewidth]{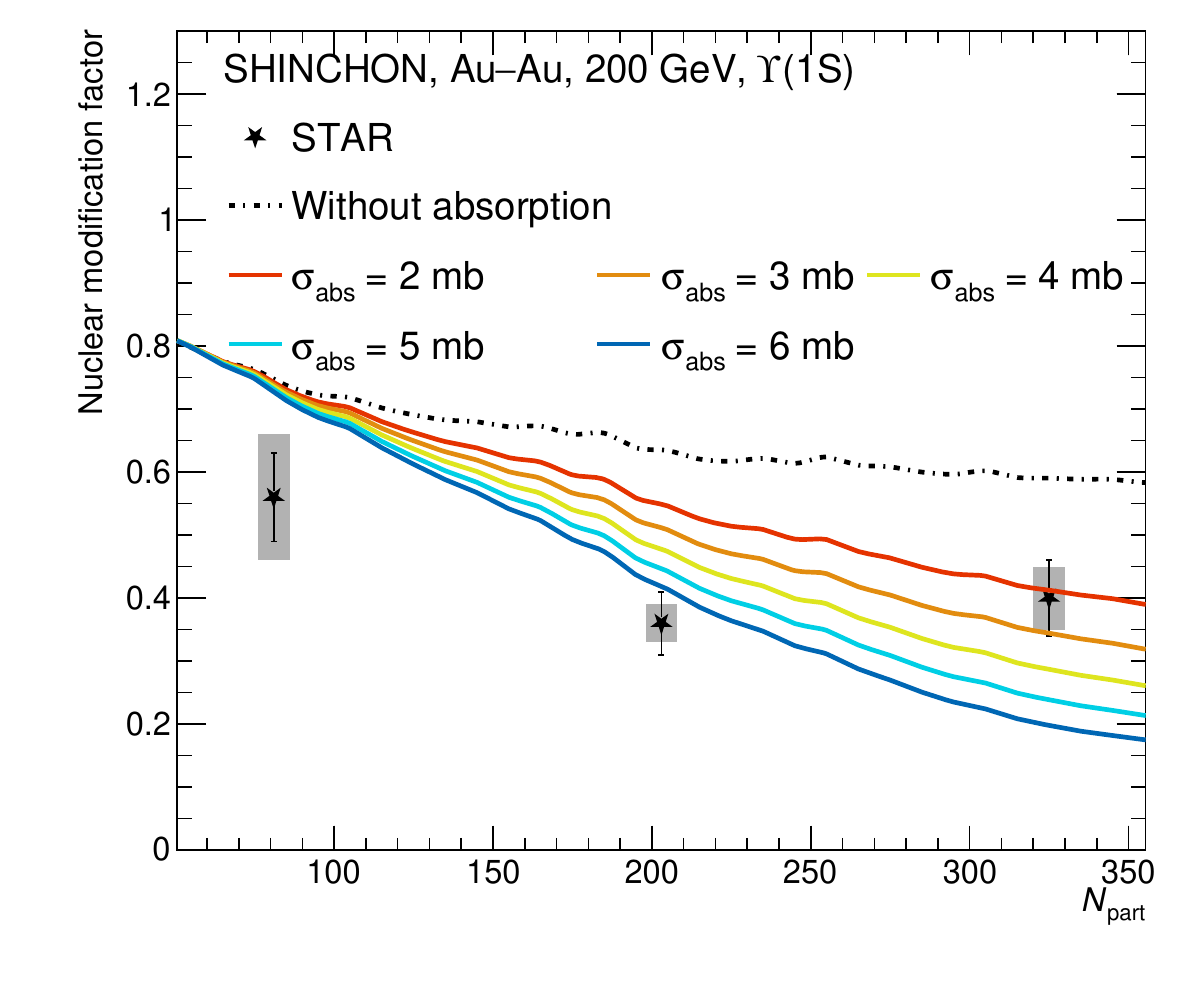}
    \caption{The nuclear modification factor of \Ups 1 as a function of participants in Au--Au collisions at $\sqsn=200$ GeV, including the nuclear absorption effect.}
    \label{fig:NMF_abs_1S}
\end{figure}

\begin{figure}[!htb]
    \centering       
    \includegraphics[width=0.99\linewidth]{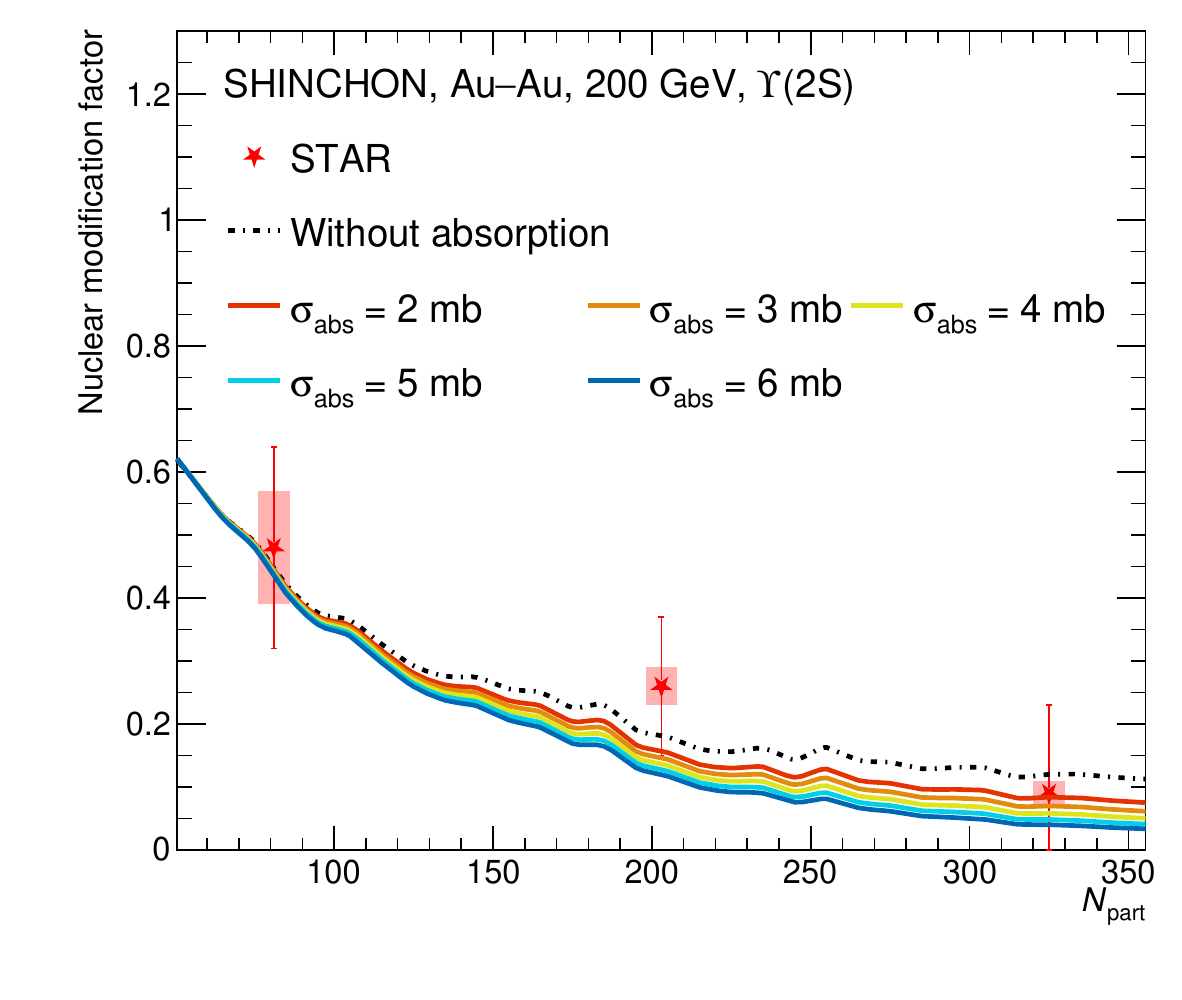}
    \caption{The nuclear modification factor of \Ups 2 as a function of participants in Au--Au collisions at $\sqsn=200$ GeV, including the nuclear absorption effect.}
    \label{fig:NMF_abs_2S}
\end{figure}

Figure~\ref{fig:NMF_abs_1S} shows the nuclear modification factor of \Ups 1 as a function of the number of participants in Au--Au collisions at $\sqsn=200$~GeV. The dash-dotted line represents the model calculation without the nuclear absorption effect, and the solid lines represent results considering the effect with different \sigabs. The model calculation, including an additional suppression effect due to the nuclear absorption, provides a better description of the experimental data. We also apply the nuclear absorption effect for \Ups 2 in Au--Au collisions $\sqsn=200$~GeV as presented in Fig.~\ref{fig:NMF_abs_2S}. We use $\tform=1.2~\mathrm{fm}/c$ for \Ups 2, which agrees better with the experimental data shown in Fig.~\ref{fig:NMF_tform_AuAu}. In the case of \Ups 2, the experimental data tends to favor the model calculations without the nuclear absorption. One possible explanation for the smaller nuclear absorption effect for \Ups 2 compared to \Ups 1 is a longer \tform of \Ups 2 relative to the beam crossing time. However, it is difficult to compare quantitatively due to the large uncertainties of the experimental data. 



\section{Summary}
\label{sec:Sum}
Based on the SHINCHON framework for the medium response of $\Upsilon$ described in Ref.~\cite{Kim:2022lgu}, we have extended the model study for heavy-ion collisions at different collision energies, such as Pb-Pb collisions at $\sqsn=5.02$~TeV and Au--Au collisions at $\sqsn=200$~GeV. In this study, we utilize the \trento model to generate initial collision geometry to evaluate the effect of different initial collision geometry on the results of $\Upsilon$ modification at the end. The initial temperature profiles for the hydrodynamic simulation have been constructed using \trento with a different reduced thickness parameter ($p$) and converted to energy density to possess the same final-state charged-particle multiplicity as experimental data in both systems. In addition, we investigate how the magnitude of medium response evolves with the formation time of quarkonia by varying the formation time in the simulation.

The nuclear modification factor from the model shows a decreasing trend with an increasing number of participants and a sequential ordering of suppression in both collision systems and energies. There is a small difference in the results using different reduced thickness parameters in the \trento model, but it is less significant than the uncertainties of experimental measurements. In Pb--Pb collisions at $\sqsn=5.02$~TeV, the model calculation with the default formation time parameters agrees with the \Ups 1 data but slightly overestimates the suppression for excited states. The difference is covered by the results of various formation times. However, even with different formation times, the model calculation does not describe the \Ups 1 data in Au--Au collisions at $\sqsn=200$~GeV, showing a similar suppression with the data at 5.02~TeV. 

To explain the different nuclear modification factors of \Ups 1 at 200~GeV, we explore the nuclear absorption effect, which can provide a better description of the suppression of \jpsi and $\Upsilon$ at backward rapidity in $p/d$--Au collisions at 200~GeV. When including the nuclear absorption effect with absorption cross-section from 2~mb to 6~mb, the model calculation shows a better agreement with the \Ups 1 data but slightly overestimates the suppression of \Ups 2. This could be related to the longer formation time of \Ups 2 compared to the beam crossing time. Experimental measurements of \Ups n in $p$--$A$ and $A$--$A$ at different collision energies can help further understand the relation between quarkonia's formation time and their modification from final-state effects.

\section{Acknowledgments}
The work was supported by the National Research Foundation of Korea (NRF) grant funded by the Korean government (MSIT) under Project No. 2018R1A5A1025563, 2020R1C1C1004985, and NRF-2008-00458. We also acknowledge technical support from KIAF administrators at KISTI.


\clearpage

\newpage
\bibliography{main}

\begin{thebibliography}{42}%
\makeatletter
\providecommand \@ifxundefined [1]{%
 \@ifx{#1\undefined}
}%
\providecommand \@ifnum [1]{%
 \ifnum #1\expandafter \@firstoftwo
 \else \expandafter \@secondoftwo
 \fi
}%
\providecommand \@ifx [1]{%
 \ifx #1\expandafter \@firstoftwo
 \else \expandafter \@secondoftwo
 \fi
}%
\providecommand \natexlab [1]{#1}%
\providecommand \enquote  [1]{``#1''}%
\providecommand \bibnamefont  [1]{#1}%
\providecommand \bibfnamefont [1]{#1}%
\providecommand \citenamefont [1]{#1}%
\providecommand \href@noop [0]{\@secondoftwo}%
\providecommand \href [0]{\begingroup \@sanitize@url \@href}%
\providecommand \@href[1]{\@@startlink{#1}\@@href}%
\providecommand \@@href[1]{\endgroup#1\@@endlink}%
\providecommand \@sanitize@url [0]{\catcode `\\12\catcode `\$12\catcode `\&12\catcode `\#12\catcode `\^12\catcode `\_12\catcode `\%12\relax}%
\providecommand \@@startlink[1]{}%
\providecommand \@@endlink[0]{}%
\providecommand \url  [0]{\begingroup\@sanitize@url \@url }%
\providecommand \@url [1]{\endgroup\@href {#1}{\urlprefix }}%
\providecommand \urlprefix  [0]{URL }%
\providecommand \Eprint [0]{\href }%
\providecommand \doibase [0]{http://dx.doi.org/}%
\providecommand \selectlanguage [0]{\@gobble}%
\providecommand \bibinfo  [0]{\@secondoftwo}%
\providecommand \bibfield  [0]{\@secondoftwo}%
\providecommand \translation [1]{[#1]}%
\providecommand \BibitemOpen [0]{}%
\providecommand \bibitemStop [0]{}%
\providecommand \bibitemNoStop [0]{.\EOS\space}%
\providecommand \EOS [0]{\spacefactor3000\relax}%
\providecommand \BibitemShut  [1]{\csname bibitem#1\endcsname}%
\let\auto@bib@innerbib\@empty
\bibitem [{\citenamefont {Busza}\ \emph {et~al.}(2018)\citenamefont {Busza}, \citenamefont {Rajagopal},\ and\ \citenamefont {van~der Schee}}]{Busza:2018rrf}%
  \BibitemOpen
  \bibfield  {author} {\bibinfo {author} {\bibfnamefont {W.}~\bibnamefont {Busza}}, \bibinfo {author} {\bibfnamefont {K.}~\bibnamefont {Rajagopal}}, \ and\ \bibinfo {author} {\bibfnamefont {W.}~\bibnamefont {van~der Schee}},\ }\bibfield  {title} {\enquote {\bibinfo {title} {{Heavy Ion Collisions: The Big Picture, and the Big Questions}},}\ }\href {\doibase 10.1146/annurev-nucl-101917-020852} {\bibfield  {journal} {\bibinfo  {journal} {Ann. Rev. Nucl. Part. Sci.}\ }\textbf {\bibinfo {volume} {68}},\ \bibinfo {pages} {339} (\bibinfo {year} {2018})},\ \Eprint {http://arxiv.org/abs/1802.04801} {arXiv:1802.04801 [hep-ph]} \BibitemShut {NoStop}%
\bibitem [{\citenamefont {Karsch}\ \emph {et~al.}(2000)\citenamefont {Karsch}, \citenamefont {Laermann},\ and\ \citenamefont {Peikert}}]{Karsch:2000ps}%
  \BibitemOpen
  \bibfield  {author} {\bibinfo {author} {\bibfnamefont {F.}~\bibnamefont {Karsch}}, \bibinfo {author} {\bibfnamefont {E.}~\bibnamefont {Laermann}}, \ and\ \bibinfo {author} {\bibfnamefont {A.}~\bibnamefont {Peikert}},\ }\bibfield  {title} {\enquote {\bibinfo {title} {{The Pressure in two flavor, (2+1)-flavor and three flavor QCD}},}\ }\href {\doibase 10.1016/S0370-2693(00)00292-6} {\bibfield  {journal} {\bibinfo  {journal} {Phys. Lett. B}\ }\textbf {\bibinfo {volume} {478}},\ \bibinfo {pages} {447--455} (\bibinfo {year} {2000})},\ \Eprint {http://arxiv.org/abs/hep-lat/0002003} {arXiv:hep-lat/0002003} \BibitemShut {NoStop}%
\bibitem [{\citenamefont {Shuryak}(1978)}]{Shuryak:1977ut}%
  \BibitemOpen
  \bibfield  {author} {\bibinfo {author} {\bibfnamefont {Edward~V.}\ \bibnamefont {Shuryak}},\ }\bibfield  {title} {\enquote {\bibinfo {title} {{Theory of Hadronic Plasma}},}\ }\href@noop {} {\bibfield  {journal} {\bibinfo  {journal} {Sov. Phys. JETP}\ }\textbf {\bibinfo {volume} {47}},\ \bibinfo {pages} {212--219} (\bibinfo {year} {1978})},\ \bibinfo {note} {[Zh. Eksp. Teor. Fiz.74,408(1978)]}\BibitemShut {NoStop}%
\bibitem [{\citenamefont {Matsui}\ and\ \citenamefont {Satz}(1986)}]{Matsui:1986dk}%
  \BibitemOpen
  \bibfield  {author} {\bibinfo {author} {\bibfnamefont {T.}~\bibnamefont {Matsui}}\ and\ \bibinfo {author} {\bibfnamefont {H.}~\bibnamefont {Satz}},\ }\bibfield  {title} {\enquote {\bibinfo {title} {\jpsi suppression by quark-gluon plasma formation},}\ }\href {\doibase 10.1016/0370-2693(86)91404-8} {\bibfield  {journal} {\bibinfo  {journal} {Phys. Lett. B}\ }\textbf {\bibinfo {volume} {178}},\ \bibinfo {pages} {416} (\bibinfo {year} {1986})}\BibitemShut {NoStop}%
\bibitem [{\citenamefont {Digal}\ \emph {et~al.}(2001)\citenamefont {Digal}, \citenamefont {Petreczky},\ and\ \citenamefont {Satz}}]{Digal:2001ue}%
  \BibitemOpen
  \bibfield  {author} {\bibinfo {author} {\bibfnamefont {S.}~\bibnamefont {Digal}}, \bibinfo {author} {\bibfnamefont {P.}~\bibnamefont {Petreczky}}, \ and\ \bibinfo {author} {\bibfnamefont {H.}~\bibnamefont {Satz}},\ }\bibfield  {title} {\enquote {\bibinfo {title} {{Quarkonium feed down and sequential suppression}},}\ }\href {\doibase 10.1103/PhysRevD.64.094015} {\bibfield  {journal} {\bibinfo  {journal} {Phys. Rev. D}\ }\textbf {\bibinfo {volume} {64}},\ \bibinfo {pages} {094015} (\bibinfo {year} {2001})},\ \Eprint {http://arxiv.org/abs/hep-ph/0106017} {arXiv:hep-ph/0106017} \BibitemShut {NoStop}%
\bibitem [{\citenamefont {Laine}\ \emph {et~al.}(2007)\citenamefont {Laine}, \citenamefont {Philipsen}, \citenamefont {Romatschke},\ and\ \citenamefont {Tassler}}]{laine:2007}%
  \BibitemOpen
  \bibfield  {author} {\bibinfo {author} {\bibfnamefont {M.}~\bibnamefont {Laine}}, \bibinfo {author} {\bibfnamefont {O.}~\bibnamefont {Philipsen}}, \bibinfo {author} {\bibfnamefont {P.}~\bibnamefont {Romatschke}}, \ and\ \bibinfo {author} {\bibfnamefont {M.}~\bibnamefont {Tassler}},\ }\bibfield  {title} {\enquote {\bibinfo {title} {Real-time static potential in hot {QCD}},}\ }\href {\doibase 10.1088/1126-6708/2007/03/054} {\bibfield  {journal} {\bibinfo  {journal} {JHEP}\ }\textbf {\bibinfo {volume} {03}},\ \bibinfo {pages} {054} (\bibinfo {year} {2007})},\ \Eprint {http://arxiv.org/abs/hep-ph/0611300} {arXiv:hep-ph/0611300 [hep-ph]} \BibitemShut {NoStop}%
\bibitem [{\citenamefont {Brambilla}\ \emph {et~al.}(2008)\citenamefont {Brambilla}, \citenamefont {Ghiglieri}, \citenamefont {Vairo},\ and\ \citenamefont {Petreczky}}]{Brambilla:2008cx}%
  \BibitemOpen
  \bibfield  {author} {\bibinfo {author} {\bibfnamefont {Nora}\ \bibnamefont {Brambilla}}, \bibinfo {author} {\bibfnamefont {Jacopo}\ \bibnamefont {Ghiglieri}}, \bibinfo {author} {\bibfnamefont {Antonio}\ \bibnamefont {Vairo}}, \ and\ \bibinfo {author} {\bibfnamefont {Peter}\ \bibnamefont {Petreczky}},\ }\bibfield  {title} {\enquote {\bibinfo {title} {Static quark antiquark pairs at finite temperature},}\ }\href {\doibase 10.1103/PhysRevD.78.014017} {\bibfield  {journal} {\bibinfo  {journal} {Phys. Rev. D}\ }\textbf {\bibinfo {volume} {78}},\ \bibinfo {pages} {014017} (\bibinfo {year} {2008})},\ \Eprint {http://arxiv.org/abs/0804.0993} {arXiv:0804.0993 [hep-ph]} \BibitemShut {NoStop}%
\bibitem [{\citenamefont {Brambilla}\ \emph {et~al.}(2010)\citenamefont {Brambilla}, \citenamefont {Escobedo}, \citenamefont {Ghiglieri}, \citenamefont {Soto},\ and\ \citenamefont {Vairo}}]{Brambilla:2010vq}%
  \BibitemOpen
  \bibfield  {author} {\bibinfo {author} {\bibfnamefont {Nora}\ \bibnamefont {Brambilla}}, \bibinfo {author} {\bibfnamefont {Miguel~Angel}\ \bibnamefont {Escobedo}}, \bibinfo {author} {\bibfnamefont {Jacopo}\ \bibnamefont {Ghiglieri}}, \bibinfo {author} {\bibfnamefont {Joan}\ \bibnamefont {Soto}}, \ and\ \bibinfo {author} {\bibfnamefont {Antonio}\ \bibnamefont {Vairo}},\ }\bibfield  {title} {\enquote {\bibinfo {title} {Heavy quarkonium in a weakly coupled quark-gluon plasma below the melting temperature},}\ }\href {\doibase 10.1007/JHEP09(2010)038} {\bibfield  {journal} {\bibinfo  {journal} {JHEP}\ }\textbf {\bibinfo {volume} {09}},\ \bibinfo {pages} {038} (\bibinfo {year} {2010})},\ \Eprint {http://arxiv.org/abs/1007.4156} {arXiv:1007.4156 [hep-ph]} \BibitemShut {NoStop}%
\bibitem [{\citenamefont {Gorenstein}\ \emph {et~al.}(2001)\citenamefont {Gorenstein}, \citenamefont {Kostyuk}, \citenamefont {Stoecker},\ and\ \citenamefont {Greiner}}]{Gorenstein:2000ck}%
  \BibitemOpen
  \bibfield  {author} {\bibinfo {author} {\bibfnamefont {Mark~I.}\ \bibnamefont {Gorenstein}}, \bibinfo {author} {\bibfnamefont {A.~P.}\ \bibnamefont {Kostyuk}}, \bibinfo {author} {\bibfnamefont {Horst}\ \bibnamefont {Stoecker}}, \ and\ \bibinfo {author} {\bibfnamefont {W.}~\bibnamefont {Greiner}},\ }\bibfield  {title} {\enquote {\bibinfo {title} {{Statistical coalescence model with exact charm conservation}},}\ }\href {\doibase 10.1016/S0370-2693(01)00516-0} {\bibfield  {journal} {\bibinfo  {journal} {Phys. Lett. B}\ }\textbf {\bibinfo {volume} {509}},\ \bibinfo {pages} {277} (\bibinfo {year} {2001})},\ \Eprint {http://arxiv.org/abs/hep-ph/0010148} {arXiv:hep-ph/0010148 [hep-ph]} \BibitemShut {NoStop}%
\bibitem [{\citenamefont {Andronic}\ \emph {et~al.}(2007)\citenamefont {Andronic}, \citenamefont {Braun-Munzinger}, \citenamefont {Redlich},\ and\ \citenamefont {Stachel}}]{Andronic:2007bi}%
  \BibitemOpen
  \bibfield  {author} {\bibinfo {author} {\bibfnamefont {A.}~\bibnamefont {Andronic}}, \bibinfo {author} {\bibfnamefont {P.}~\bibnamefont {Braun-Munzinger}}, \bibinfo {author} {\bibfnamefont {K.}~\bibnamefont {Redlich}}, \ and\ \bibinfo {author} {\bibfnamefont {J.}~\bibnamefont {Stachel}},\ }\bibfield  {title} {\enquote {\bibinfo {title} {Evidence for charmonium generation at the phase boundary in ultra-relativistic nuclear collisions},}\ }\href {\doibase 10.1016/j.physletb.2007.07.036} {\bibfield  {journal} {\bibinfo  {journal} {Phys. Lett. B}\ }\textbf {\bibinfo {volume} {652}},\ \bibinfo {pages} {259} (\bibinfo {year} {2007})},\ \Eprint {http://arxiv.org/abs/nucl-th/0701079} {arXiv:nucl-th/0701079 [NUCL-TH]} \BibitemShut {NoStop}%
\bibitem [{\citenamefont {Ravagli}\ and\ \citenamefont {Rapp}(2007)}]{Ravagli:2007xx}%
  \BibitemOpen
  \bibfield  {author} {\bibinfo {author} {\bibfnamefont {L.}~\bibnamefont {Ravagli}}\ and\ \bibinfo {author} {\bibfnamefont {R.}~\bibnamefont {Rapp}},\ }\bibfield  {title} {\enquote {\bibinfo {title} {{Quark Coalescence based on a Transport Equation}},}\ }\href {\doibase 10.1016/j.physletb.2007.07.043} {\bibfield  {journal} {\bibinfo  {journal} {Phys. Lett. B}\ }\textbf {\bibinfo {volume} {655}},\ \bibinfo {pages} {126--131} (\bibinfo {year} {2007})},\ \Eprint {http://arxiv.org/abs/0705.0021} {arXiv:0705.0021 [hep-ph]} \BibitemShut {NoStop}%
\bibitem [{\citenamefont {Blaizot}\ \emph {et~al.}(2016)\citenamefont {Blaizot}, \citenamefont {De~Boni}, \citenamefont {Faccioli},\ and\ \citenamefont {Garberoglio}}]{blaizot:2016jp}%
  \BibitemOpen
  \bibfield  {author} {\bibinfo {author} {\bibfnamefont {Jean-Paul}\ \bibnamefont {Blaizot}}, \bibinfo {author} {\bibfnamefont {Davide}\ \bibnamefont {De~Boni}}, \bibinfo {author} {\bibfnamefont {Pietro}\ \bibnamefont {Faccioli}}, \ and\ \bibinfo {author} {\bibfnamefont {Giovanni}\ \bibnamefont {Garberoglio}},\ }\bibfield  {title} {\enquote {\bibinfo {title} {Heavy quark bound states in a quark-gluon plasma: Dissociation and recombination},}\ }\href {\doibase 10.1016/j.nuclphysa.2015.10.011} {\bibfield  {journal} {\bibinfo  {journal} {Nucl. Phys. A}\ }\textbf {\bibinfo {volume} {946}},\ \bibinfo {pages} {49} (\bibinfo {year} {2016})},\ \Eprint {http://arxiv.org/abs/1503.03857} {arXiv:1503.03857 [nucl-th]} \BibitemShut {NoStop}%
\bibitem [{\citenamefont {Andronic}\ \emph {et~al.}(2016)\citenamefont {Andronic} \emph {et~al.}}]{Andronic:2015wma}%
  \BibitemOpen
  \bibfield  {author} {\bibinfo {author} {\bibfnamefont {A.}~\bibnamefont {Andronic}} \emph {et~al.},\ }\bibfield  {title} {\enquote {\bibinfo {title} {{Heavy-flavour and quarkonium production in the LHC era: from proton\textendash{}proton to heavy-ion collisions}},}\ }\href {\doibase 10.1140/epjc/s10052-015-3819-5} {\bibfield  {journal} {\bibinfo  {journal} {Eur. Phys. J. C}\ }\textbf {\bibinfo {volume} {76}},\ \bibinfo {pages} {107} (\bibinfo {year} {2016})},\ \Eprint {http://arxiv.org/abs/1506.03981} {arXiv:1506.03981 [nucl-ex]} \BibitemShut {NoStop}%
\bibitem [{\citenamefont {Aboona}\ \emph {et~al.}(2023)\citenamefont {Aboona} \emph {et~al.}}]{STAR:2022rpk}%
  \BibitemOpen
  \bibfield  {author} {\bibinfo {author} {\bibfnamefont {Bassam}\ \bibnamefont {Aboona}} \emph {et~al.} (\bibinfo {collaboration} {STAR}),\ }\bibfield  {title} {\enquote {\bibinfo {title} {{Observation of sequential $\Upsilon$ suppression in Au+Au collisions at $\sqrt{s_{_\mathrm{NN}}}$ = 200 GeV with the STAR experiment}},}\ }\href {\doibase 10.1103/PhysRevLett.130.112301} {\bibfield  {journal} {\bibinfo  {journal} {Phys. Rev. Lett.}\ }\textbf {\bibinfo {volume} {130}},\ \bibinfo {pages} {112301} (\bibinfo {year} {2023})},\ \Eprint {http://arxiv.org/abs/2207.06568} {arXiv:2207.06568 [nucl-ex]} \BibitemShut {NoStop}%
\bibitem [{\citenamefont {Acharya}\ \emph {et~al.}(2021)\citenamefont {Acharya} \emph {et~al.}}]{ALICE:2021UpsForward}%
  \BibitemOpen
  \bibfield  {author} {\bibinfo {author} {\bibfnamefont {S.}~\bibnamefont {Acharya}} \emph {et~al.} (\bibinfo {collaboration} {ALICE}),\ }\bibfield  {title} {\enquote {\bibinfo {title} {{$\Upsilon$ production and nuclear modification at forward rapidity in Pb-Pb collisions at $\sqrt{s_{_\mathrm{NN}}} =$ = 5.02 TeV}},}\ }\href {\doibase 10.1016/j.physletb.2021.136579} {\bibfield  {journal} {\bibinfo  {journal} {Phys. Lett. B}\ }\textbf {\bibinfo {volume} {822}},\ \bibinfo {pages} {136579} (\bibinfo {year} {2021})},\ \Eprint {http://arxiv.org/abs/2011.05758} {arXiv:2011.05758 [nucl-ex]} \BibitemShut {NoStop}%
\bibitem [{ATL(2022)}]{ATLAS:2022xso}%
  \BibitemOpen
  \href@noop {} {\enquote {\bibinfo {title} {{Production of $\varUpsilon(\textrm{nS})$ mesons in Pb+Pb and $pp$ collisions at 5.02 TeV}},}\ } (\bibinfo {year} {2022}),\ \Eprint {http://arxiv.org/abs/2205.03042} {arXiv:2205.03042 [nucl-ex]} \BibitemShut {NoStop}%
\bibitem [{\citenamefont {Sirunyan}\ \emph {et~al.}(2019)\citenamefont {Sirunyan} \emph {et~al.}}]{CMS:2018zza}%
  \BibitemOpen
  \bibfield  {author} {\bibinfo {author} {\bibfnamefont {Albert~M}\ \bibnamefont {Sirunyan}} \emph {et~al.} (\bibinfo {collaboration} {CMS}),\ }\bibfield  {title} {\enquote {\bibinfo {title} {{Measurement of nuclear modification factors of $\Upsilon$(1S), $\Upsilon$(2S), and $\Upsilon$(3S) mesons in PbPb collisions at $\sqrt{s_{_\mathrm{NN}}} =$ 5.02 TeV}},}\ }\href {\doibase 10.1016/j.physletb.2019.01.006} {\bibfield  {journal} {\bibinfo  {journal} {Phys. Lett. B}\ }\textbf {\bibinfo {volume} {790}},\ \bibinfo {pages} {270} (\bibinfo {year} {2019})},\ \Eprint {http://arxiv.org/abs/1805.09215} {arXiv:1805.09215 [hep-ex]} \BibitemShut {NoStop}%
\bibitem [{\citenamefont {Tumasyan}\ \emph {et~al.}(2023)\citenamefont {Tumasyan} \emph {et~al.}}]{CMS:2023lfu}%
  \BibitemOpen
  \bibfield  {author} {\bibinfo {author} {\bibfnamefont {Armen}\ \bibnamefont {Tumasyan}} \emph {et~al.} (\bibinfo {collaboration} {CMS}),\ }\bibfield  {title} {\enquote {\bibinfo {title} {{Observation of the $\Upsilon$(3S) meson and suppression of $\Upsilon$ states in PbPb collisions at $\sqrt{s_\mathrm{NN}}$ = 5.02 TeV}},}\ }\href@noop {} {\  (\bibinfo {year} {2023})},\ \Eprint {http://arxiv.org/abs/2303.17026} {arXiv:2303.17026 [hep-ex]} \BibitemShut {NoStop}%
\bibitem [{\citenamefont {Hong}\ and\ \citenamefont {Lee}(2020)}]{Hong:2019ade}%
  \BibitemOpen
  \bibfield  {author} {\bibinfo {author} {\bibfnamefont {Juhee}\ \bibnamefont {Hong}}\ and\ \bibinfo {author} {\bibfnamefont {Su~Houng}\ \bibnamefont {Lee}},\ }\bibfield  {title} {\enquote {\bibinfo {title} {{$\Upsilon(1S)$ transverse momentum spectra through dissociation and regeneration in heavy-ion collisions}},}\ }\href {\doibase 10.1016/j.physletb.2019.135147} {\bibfield  {journal} {\bibinfo  {journal} {Phys. Lett. B}\ }\textbf {\bibinfo {volume} {801}},\ \bibinfo {pages} {135147} (\bibinfo {year} {2020})},\ \Eprint {http://arxiv.org/abs/1909.07696} {arXiv:1909.07696 [nucl-th]} \BibitemShut {NoStop}%
\bibitem [{\citenamefont {Rothkopf}\ \emph {et~al.}(2012)\citenamefont {Rothkopf}, \citenamefont {Hatsuda},\ and\ \citenamefont {Sasaki}}]{Rothkopf:2011db}%
  \BibitemOpen
  \bibfield  {author} {\bibinfo {author} {\bibfnamefont {Alexander}\ \bibnamefont {Rothkopf}}, \bibinfo {author} {\bibfnamefont {Tetsuo}\ \bibnamefont {Hatsuda}}, \ and\ \bibinfo {author} {\bibfnamefont {Shoichi}\ \bibnamefont {Sasaki}},\ }\bibfield  {title} {\enquote {\bibinfo {title} {{Complex Heavy-Quark Potential at Finite Temperature from Lattice QCD}},}\ }\href {\doibase 10.1103/PhysRevLett.108.162001} {\bibfield  {journal} {\bibinfo  {journal} {Phys. Rev. Lett.}\ }\textbf {\bibinfo {volume} {108}},\ \bibinfo {pages} {162001} (\bibinfo {year} {2012})},\ \Eprint {http://arxiv.org/abs/1108.1579} {arXiv:1108.1579 [hep-lat]} \BibitemShut {NoStop}%
\bibitem [{\citenamefont {Lafferty}\ and\ \citenamefont {Rothkopf}(2020)}]{Lafferty:2019jpr}%
  \BibitemOpen
  \bibfield  {author} {\bibinfo {author} {\bibfnamefont {David}\ \bibnamefont {Lafferty}}\ and\ \bibinfo {author} {\bibfnamefont {Alexander}\ \bibnamefont {Rothkopf}},\ }\bibfield  {title} {\enquote {\bibinfo {title} {{Improved Gauss law model and in-medium heavy quarkonium at finite density and velocity}},}\ }\href {\doibase 10.1103/PhysRevD.101.056010} {\bibfield  {journal} {\bibinfo  {journal} {Phys. Rev. D}\ }\textbf {\bibinfo {volume} {101}},\ \bibinfo {pages} {056010} (\bibinfo {year} {2020})},\ \Eprint {http://arxiv.org/abs/1906.00035} {arXiv:1906.00035 [hep-ph]} \BibitemShut {NoStop}%
\bibitem [{\citenamefont {Liu}\ and\ \citenamefont {Rapp}(2018)}]{Liu:2017qah}%
  \BibitemOpen
  \bibfield  {author} {\bibinfo {author} {\bibfnamefont {Shuai Y.~F.}\ \bibnamefont {Liu}}\ and\ \bibinfo {author} {\bibfnamefont {Ralf}\ \bibnamefont {Rapp}},\ }\bibfield  {title} {\enquote {\bibinfo {title} {{$T$-matrix Approach to Quark-Gluon Plasma}},}\ }\href {\doibase 10.1103/PhysRevC.97.034918} {\bibfield  {journal} {\bibinfo  {journal} {Phys. Rev. C}\ }\textbf {\bibinfo {volume} {97}},\ \bibinfo {pages} {034918} (\bibinfo {year} {2018})},\ \Eprint {http://arxiv.org/abs/1711.03282} {arXiv:1711.03282 [nucl-th]} \BibitemShut {NoStop}%
\bibitem [{\citenamefont {Brambilla}\ \emph {et~al.}(2022)\citenamefont {Brambilla}, \citenamefont {Escobedo}, \citenamefont {Islam}, \citenamefont {Strickland}, \citenamefont {Tiwari}, \citenamefont {Vairo},\ and\ \citenamefont {Vander~Griend}}]{Brambilla:2022ynh}%
  \BibitemOpen
  \bibfield  {author} {\bibinfo {author} {\bibfnamefont {Nora}\ \bibnamefont {Brambilla}}, \bibinfo {author} {\bibfnamefont {Miguel~\'Angel}\ \bibnamefont {Escobedo}}, \bibinfo {author} {\bibfnamefont {Ajaharul}\ \bibnamefont {Islam}}, \bibinfo {author} {\bibfnamefont {Michael}\ \bibnamefont {Strickland}}, \bibinfo {author} {\bibfnamefont {Anurag}\ \bibnamefont {Tiwari}}, \bibinfo {author} {\bibfnamefont {Antonio}\ \bibnamefont {Vairo}}, \ and\ \bibinfo {author} {\bibfnamefont {Peter}\ \bibnamefont {Vander~Griend}},\ }\bibfield  {title} {\enquote {\bibinfo {title} {{Heavy quarkonium dynamics at next-to-leading order in the binding energy over temperature}},}\ }\href {\doibase 10.1007/JHEP08(2022)303} {\bibfield  {journal} {\bibinfo  {journal} {JHEP}\ }\textbf {\bibinfo {volume} {08}},\ \bibinfo {pages} {303} (\bibinfo {year} {2022})},\ \Eprint {http://arxiv.org/abs/2205.10289} {arXiv:2205.10289 [hep-ph]} \BibitemShut {NoStop}%
\bibitem [{\citenamefont {Du}\ \emph {et~al.}(2017)\citenamefont {Du}, \citenamefont {He},\ and\ \citenamefont {Rapp}}]{Du:2017qkv}%
  \BibitemOpen
  \bibfield  {author} {\bibinfo {author} {\bibfnamefont {Xiaojian}\ \bibnamefont {Du}}, \bibinfo {author} {\bibfnamefont {Min}\ \bibnamefont {He}}, \ and\ \bibinfo {author} {\bibfnamefont {Ralf}\ \bibnamefont {Rapp}},\ }\bibfield  {title} {\enquote {\bibinfo {title} {{Color Screening and Regeneration of Bottomonia in High-Energy Heavy-Ion Collisions}},}\ }\href {\doibase 10.1103/PhysRevC.96.054901} {\bibfield  {journal} {\bibinfo  {journal} {Phys. Rev. C}\ }\textbf {\bibinfo {volume} {96}},\ \bibinfo {pages} {054901} (\bibinfo {year} {2017})},\ \Eprint {http://arxiv.org/abs/1706.08670} {arXiv:1706.08670 [hep-ph]} \BibitemShut {NoStop}%
\bibitem [{\citenamefont {Kim}\ \emph {et~al.}(2023)\citenamefont {Kim}, \citenamefont {Seo}, \citenamefont {Hong}, \citenamefont {Hong}, \citenamefont {Kim}, \citenamefont {Kim}, \citenamefont {Kweon}, \citenamefont {Lee}, \citenamefont {Lim},\ and\ \citenamefont {Park}}]{Kim:2022lgu}%
  \BibitemOpen
  \bibfield  {author} {\bibinfo {author} {\bibfnamefont {Junlee}\ \bibnamefont {Kim}}, \bibinfo {author} {\bibfnamefont {Jinjoo}\ \bibnamefont {Seo}}, \bibinfo {author} {\bibfnamefont {Byungsik}\ \bibnamefont {Hong}}, \bibinfo {author} {\bibfnamefont {Juhee}\ \bibnamefont {Hong}}, \bibinfo {author} {\bibfnamefont {Eun-Joo}\ \bibnamefont {Kim}}, \bibinfo {author} {\bibfnamefont {Yongsun}\ \bibnamefont {Kim}}, \bibinfo {author} {\bibfnamefont {MinJung}\ \bibnamefont {Kweon}}, \bibinfo {author} {\bibfnamefont {Su~Houng}\ \bibnamefont {Lee}}, \bibinfo {author} {\bibfnamefont {Sanghoon}\ \bibnamefont {Lim}}, \ and\ \bibinfo {author} {\bibfnamefont {Jaebeom}\ \bibnamefont {Park}},\ }\bibfield  {title} {\enquote {\bibinfo {title} {{Model study on \ensuremath{\Upsilon}(nS) modification in small collision systems}},}\ }\href {\doibase 10.1103/PhysRevC.107.054905} {\bibfield  {journal} {\bibinfo  {journal} {Phys. Rev. C}\ }\textbf {\bibinfo {volume} {107}},\ \bibinfo {pages} {054905} (\bibinfo {year} {2023})},\ \Eprint
  {http://arxiv.org/abs/2209.12303} {arXiv:2209.12303 [nucl-th]} \BibitemShut {NoStop}%
\bibitem [{\citenamefont {Moreland}\ \emph {et~al.}(2015)\citenamefont {Moreland}, \citenamefont {Bernhard},\ and\ \citenamefont {Bass}}]{Moreland:2014oya}%
  \BibitemOpen
  \bibfield  {author} {\bibinfo {author} {\bibfnamefont {J.~Scott}\ \bibnamefont {Moreland}}, \bibinfo {author} {\bibfnamefont {Jonah~E.}\ \bibnamefont {Bernhard}}, \ and\ \bibinfo {author} {\bibfnamefont {Steffen~A.}\ \bibnamefont {Bass}},\ }\bibfield  {title} {\enquote {\bibinfo {title} {{Alternative ansatz to wounded nucleon and binary collision scaling in high-energy nuclear collisions}},}\ }\href {\doibase 10.1103/PhysRevC.92.011901} {\bibfield  {journal} {\bibinfo  {journal} {Phys. Rev. C}\ }\textbf {\bibinfo {volume} {92}},\ \bibinfo {pages} {011901} (\bibinfo {year} {2015})},\ \Eprint {http://arxiv.org/abs/1412.4708} {arXiv:1412.4708 [nucl-th]} \BibitemShut {NoStop}%
\bibitem [{\citenamefont {Romatschke}(2015)}]{Romatschke:2015gxa}%
  \BibitemOpen
  \bibfield  {author} {\bibinfo {author} {\bibfnamefont {Paul}\ \bibnamefont {Romatschke}},\ }\bibfield  {title} {\enquote {\bibinfo {title} {{Light-Heavy Ion Collisions: A window into pre-equilibrium QCD dynamics?}}}\ }\href {\doibase 10.1140/epjc/s10052-015-3509-3} {\bibfield  {journal} {\bibinfo  {journal} {Eur. Phys. J. C}\ }\textbf {\bibinfo {volume} {75}},\ \bibinfo {pages} {305} (\bibinfo {year} {2015})},\ \Eprint {http://arxiv.org/abs/1502.04745} {arXiv:1502.04745 [nucl-th]} \BibitemShut {NoStop}%
\bibitem [{\citenamefont {McGlinchey}\ \emph {et~al.}(2013)\citenamefont {McGlinchey}, \citenamefont {Frawley},\ and\ \citenamefont {Vogt}}]{McGlinchey:2012bp}%
  \BibitemOpen
  \bibfield  {author} {\bibinfo {author} {\bibfnamefont {D.~C.}\ \bibnamefont {McGlinchey}}, \bibinfo {author} {\bibfnamefont {A.~D.}\ \bibnamefont {Frawley}}, \ and\ \bibinfo {author} {\bibfnamefont {R.}~\bibnamefont {Vogt}},\ }\bibfield  {title} {\enquote {\bibinfo {title} {{Impact parameter dependence of the nuclear modification of $J/\psi$ production in $d+$Au collisions at $\sqrt{s_{NN}} = 200$ GeV}},}\ }\href {\doibase 10.1103/PhysRevC.87.054910} {\bibfield  {journal} {\bibinfo  {journal} {Phys. Rev. C}\ }\textbf {\bibinfo {volume} {87}},\ \bibinfo {pages} {054910} (\bibinfo {year} {2013})},\ \Eprint {http://arxiv.org/abs/1208.2667} {arXiv:1208.2667 [nucl-th]} \BibitemShut {NoStop}%
\bibitem [{\citenamefont {Miller}\ \emph {et~al.}(2007)\citenamefont {Miller}, \citenamefont {Reygers}, \citenamefont {Sanders},\ and\ \citenamefont {Steinberg}}]{Miller:2007ri}%
  \BibitemOpen
  \bibfield  {author} {\bibinfo {author} {\bibfnamefont {Michael~L.}\ \bibnamefont {Miller}}, \bibinfo {author} {\bibfnamefont {Klaus}\ \bibnamefont {Reygers}}, \bibinfo {author} {\bibfnamefont {Stephen~J.}\ \bibnamefont {Sanders}}, \ and\ \bibinfo {author} {\bibfnamefont {Peter}\ \bibnamefont {Steinberg}},\ }\bibfield  {title} {\enquote {\bibinfo {title} {{Glauber modeling in high energy nuclear collisions}},}\ }\href {\doibase 10.1146/annurev.nucl.57.090506.123020} {\bibfield  {journal} {\bibinfo  {journal} {Ann. Rev. Nucl. Part. Sci.}\ }\textbf {\bibinfo {volume} {57}},\ \bibinfo {pages} {205--243} (\bibinfo {year} {2007})},\ \Eprint {http://arxiv.org/abs/nucl-ex/0701025} {arXiv:nucl-ex/0701025} \BibitemShut {NoStop}%
\bibitem [{\citenamefont {Alver}\ \emph {et~al.}(2011)\citenamefont {Alver} \emph {et~al.}}]{PHOBOS:2010eyu}%
  \BibitemOpen
  \bibfield  {author} {\bibinfo {author} {\bibfnamefont {B.}~\bibnamefont {Alver}} \emph {et~al.} (\bibinfo {collaboration} {PHOBOS}),\ }\bibfield  {title} {\enquote {\bibinfo {title} {{Phobos results on charged particle multiplicity and pseudorapidity distributions in Au+Au, Cu+Cu, d+Au, and p+p collisions at ultra-relativistic energies}},}\ }\href {\doibase 10.1103/PhysRevC.83.024913} {\bibfield  {journal} {\bibinfo  {journal} {Phys. Rev. C}\ }\textbf {\bibinfo {volume} {83}},\ \bibinfo {pages} {024913} (\bibinfo {year} {2011})},\ \Eprint {http://arxiv.org/abs/1011.1940} {arXiv:1011.1940 [nucl-ex]} \BibitemShut {NoStop}%
\bibitem [{\citenamefont {Adam}\ \emph {et~al.}(2016)\citenamefont {Adam} \emph {et~al.}}]{ALICE:2015juo}%
  \BibitemOpen
  \bibfield  {author} {\bibinfo {author} {\bibfnamefont {Jaroslav}\ \bibnamefont {Adam}} \emph {et~al.} (\bibinfo {collaboration} {ALICE}),\ }\bibfield  {title} {\enquote {\bibinfo {title} {{Centrality Dependence of the Charged-Particle Multiplicity Density at Midrapidity in Pb--Pb Collisions at $\sqrt{s_{\rm NN}}$ = 5.02 TeV}},}\ }\href {\doibase 10.1103/PhysRevLett.116.222302} {\bibfield  {journal} {\bibinfo  {journal} {Phys. Rev. Lett.}\ }\textbf {\bibinfo {volume} {116}},\ \bibinfo {pages} {222302} (\bibinfo {year} {2016})},\ \Eprint {http://arxiv.org/abs/1512.06104} {arXiv:1512.06104 [nucl-ex]} \BibitemShut {NoStop}%
\bibitem [{\citenamefont {Acharya}\ \emph {et~al.}(2019)\citenamefont {Acharya} \emph {et~al.}}]{ALICE:2018cpu}%
  \BibitemOpen
  \bibfield  {author} {\bibinfo {author} {\bibfnamefont {Shreyasi}\ \bibnamefont {Acharya}} \emph {et~al.} (\bibinfo {collaboration} {ALICE}),\ }\bibfield  {title} {\enquote {\bibinfo {title} {{Centrality and pseudorapidity dependence of the charged-particle multiplicity density in Xe\textendash{}Xe collisions at $\sqrt{s_{\rm NN}}$ =5.44TeV}},}\ }\href {\doibase 10.1016/j.physletb.2018.12.048} {\bibfield  {journal} {\bibinfo  {journal} {Phys. Lett. B}\ }\textbf {\bibinfo {volume} {790}},\ \bibinfo {pages} {35--48} (\bibinfo {year} {2019})},\ \Eprint {http://arxiv.org/abs/1805.04432} {arXiv:1805.04432 [nucl-ex]} \BibitemShut {NoStop}%
\bibitem [{\citenamefont {Acharya}\ \emph {et~al.}(2018)\citenamefont {Acharya} \emph {et~al.}}]{ALICE:2018rtz}%
  \BibitemOpen
  \bibfield  {author} {\bibinfo {author} {\bibfnamefont {S.}~\bibnamefont {Acharya}} \emph {et~al.} (\bibinfo {collaboration} {ALICE}),\ }\bibfield  {title} {\enquote {\bibinfo {title} {{Energy dependence and fluctuations of anisotropic flow in Pb--Pb collisions at $ \sqrt{s_{\mathrm{NN}}}=5.02 $ and 2.76 TeV}},}\ }\href {\doibase 10.1007/JHEP07(2018)103} {\bibfield  {journal} {\bibinfo  {journal} {JHEP}\ }\textbf {\bibinfo {volume} {07}},\ \bibinfo {pages} {103} (\bibinfo {year} {2018})},\ \Eprint {http://arxiv.org/abs/1804.02944} {arXiv:1804.02944 [nucl-ex]} \BibitemShut {NoStop}%
\bibitem [{\citenamefont {Afanasiev}\ \emph {et~al.}(2009)\citenamefont {Afanasiev} \emph {et~al.}}]{PHENIX:2009cjr}%
  \BibitemOpen
  \bibfield  {author} {\bibinfo {author} {\bibfnamefont {S.}~\bibnamefont {Afanasiev}} \emph {et~al.} (\bibinfo {collaboration} {PHENIX}),\ }\bibfield  {title} {\enquote {\bibinfo {title} {{Systematic Studies of Elliptic Flow Measurements in Au+Au Collisions at $\sqrt{s_{NN}}$= 200-GeV}},}\ }\href {\doibase 10.1103/PhysRevC.80.024909} {\bibfield  {journal} {\bibinfo  {journal} {Phys. Rev. C}\ }\textbf {\bibinfo {volume} {80}},\ \bibinfo {pages} {024909} (\bibinfo {year} {2009})},\ \Eprint {http://arxiv.org/abs/0905.1070} {arXiv:0905.1070 [nucl-ex]} \BibitemShut {NoStop}%
\bibitem [{\citenamefont {Mocsy}\ \emph {et~al.}(2013)\citenamefont {Mocsy}, \citenamefont {Petreczky},\ and\ \citenamefont {Strickland}}]{MPS:2013qqgp}%
  \BibitemOpen
  \bibfield  {author} {\bibinfo {author} {\bibfnamefont {Agnes}\ \bibnamefont {Mocsy}}, \bibinfo {author} {\bibfnamefont {Peter}\ \bibnamefont {Petreczky}}, \ and\ \bibinfo {author} {\bibfnamefont {Michael}\ \bibnamefont {Strickland}},\ }\bibfield  {title} {\enquote {\bibinfo {title} {{Quarkonia in the Quark Gluon Plasma}},}\ }\href {\doibase 10.1142/S0217751X13400125} {\bibfield  {journal} {\bibinfo  {journal} {Int. J. of Mod. Phys. A}\ }\textbf {\bibinfo {volume} {28}},\ \bibinfo {pages} {1340012} (\bibinfo {year} {2013})},\ \Eprint {http://arxiv.org/abs/1302.2180} {arXiv:1302.2180 [hep-ph]} \BibitemShut {NoStop}%
\bibitem [{\citenamefont {Satz}(2006)}]{HS:2006cdqb}%
  \BibitemOpen
  \bibfield  {author} {\bibinfo {author} {\bibfnamefont {Helmut}\ \bibnamefont {Satz}},\ }\bibfield  {title} {\enquote {\bibinfo {title} {{Colour deconfinement and quarkonium binding}},}\ }\href {\doibase 10.1088/0954-3899/32/3/R01} {\bibfield  {journal} {\bibinfo  {journal} {J. Phys. G: Nucl. Part. Phys.}\ }\textbf {\bibinfo {volume} {32}},\ \bibinfo {pages} {R25} (\bibinfo {year} {2006})},\ \Eprint {http://arxiv.org/abs/hep-ph/0512217} {arXiv:hep-ph/0512217 [hep-ph]} \BibitemShut {NoStop}%
\bibitem [{\citenamefont {Ferreiro}\ \emph {et~al.}(2022)\citenamefont {Ferreiro}, \citenamefont {Fleuret},\ and\ \citenamefont {Maurice}}]{Ferreiro:2021kwk}%
  \BibitemOpen
  \bibfield  {author} {\bibinfo {author} {\bibfnamefont {E.~G.}\ \bibnamefont {Ferreiro}}, \bibinfo {author} {\bibfnamefont {F.}~\bibnamefont {Fleuret}}, \ and\ \bibinfo {author} {\bibfnamefont {E.}~\bibnamefont {Maurice}},\ }\bibfield  {title} {\enquote {\bibinfo {title} {{Towards quarkonium formation time determination}},}\ }\href {\doibase 10.1140/epjc/s10052-022-10152-z} {\bibfield  {journal} {\bibinfo  {journal} {Eur. Phys. J. C}\ }\textbf {\bibinfo {volume} {82}},\ \bibinfo {pages} {201} (\bibinfo {year} {2022})},\ \Eprint {http://arxiv.org/abs/2107.01150} {arXiv:2107.01150 [hep-ph]} \BibitemShut {NoStop}%
\bibitem [{\citenamefont {Arleo}\ \emph {et~al.}(2000)\citenamefont {Arleo}, \citenamefont {Gossiaux}, \citenamefont {Gousset},\ and\ \citenamefont {Aichelin}}]{Arleo:1999af}%
  \BibitemOpen
  \bibfield  {author} {\bibinfo {author} {\bibfnamefont {F.}~\bibnamefont {Arleo}}, \bibinfo {author} {\bibfnamefont {P.~B.}\ \bibnamefont {Gossiaux}}, \bibinfo {author} {\bibfnamefont {T.}~\bibnamefont {Gousset}}, \ and\ \bibinfo {author} {\bibfnamefont {J.}~\bibnamefont {Aichelin}},\ }\bibfield  {title} {\enquote {\bibinfo {title} {{Charmonium suppression in p-A collisions}},}\ }\href {\doibase 10.1103/PhysRevC.61.054906} {\bibfield  {journal} {\bibinfo  {journal} {Phys. Rev. C}\ }\textbf {\bibinfo {volume} {61}},\ \bibinfo {pages} {054906} (\bibinfo {year} {2000})},\ \Eprint {http://arxiv.org/abs/hep-ph/9907286} {arXiv:hep-ph/9907286} \BibitemShut {NoStop}%
\bibitem [{\citenamefont {Alessandro}\ \emph {et~al.}(2006)\citenamefont {Alessandro} \emph {et~al.}}]{NA50:2006rdp}%
  \BibitemOpen
  \bibfield  {author} {\bibinfo {author} {\bibfnamefont {B.}~\bibnamefont {Alessandro}} \emph {et~al.} (\bibinfo {collaboration} {NA50}),\ }\bibfield  {title} {\enquote {\bibinfo {title} {{J/psi and psi-prime production and their normal nuclear absorption in proton-nucleus collisions at 400-GeV}},}\ }\href {\doibase 10.1140/epjc/s10052-006-0079-4} {\bibfield  {journal} {\bibinfo  {journal} {Eur. Phys. J. C}\ }\textbf {\bibinfo {volume} {48}},\ \bibinfo {pages} {329} (\bibinfo {year} {2006})},\ \Eprint {http://arxiv.org/abs/nucl-ex/0612012} {arXiv:nucl-ex/0612012} \BibitemShut {NoStop}%
\bibitem [{\citenamefont {Adare}\ \emph {et~al.}(2011)\citenamefont {Adare} \emph {et~al.}}]{PHENIX:2010hmo}%
  \BibitemOpen
  \bibfield  {author} {\bibinfo {author} {\bibfnamefont {A.}~\bibnamefont {Adare}} \emph {et~al.} (\bibinfo {collaboration} {PHENIX}),\ }\bibfield  {title} {\enquote {\bibinfo {title} {{Cold Nuclear Matter Effects on $J/\psi$ Yields as a Function of Rapidity and Nuclear Geometry in Deuteron-Gold Collisions at $\sqrt{s_{NN}}=200$ GeV}},}\ }\href {\doibase 10.1103/PhysRevLett.107.142301} {\bibfield  {journal} {\bibinfo  {journal} {Phys. Rev. Lett.}\ }\textbf {\bibinfo {volume} {107}},\ \bibinfo {pages} {142301} (\bibinfo {year} {2011})},\ \Eprint {http://arxiv.org/abs/1010.1246} {arXiv:1010.1246 [nucl-ex]} \BibitemShut {NoStop}%
\bibitem [{\citenamefont {Adare}\ \emph {et~al.}(2013)\citenamefont {Adare} \emph {et~al.}}]{PHENIX:2012xws}%
  \BibitemOpen
  \bibfield  {author} {\bibinfo {author} {\bibfnamefont {A.}~\bibnamefont {Adare}} \emph {et~al.} (\bibinfo {collaboration} {PHENIX}),\ }\bibfield  {title} {\enquote {\bibinfo {title} {{$\Upsilon(1S+2S+3S)$ production in $d+$Au and $p+p$ collisions at $\sqrt{s_{NN}}=200$ GeV and cold-nuclear matter effects}},}\ }\href {\doibase 10.1103/PhysRevC.87.044909} {\bibfield  {journal} {\bibinfo  {journal} {Phys. Rev. C}\ }\textbf {\bibinfo {volume} {87}},\ \bibinfo {pages} {044909} (\bibinfo {year} {2013})},\ \Eprint {http://arxiv.org/abs/1211.4017} {arXiv:1211.4017 [nucl-ex]} \BibitemShut {NoStop}%
\bibitem [{\citenamefont {Adam}\ \emph {et~al.}(2015)\citenamefont {Adam} \emph {et~al.}}]{ALICE:2015kgk}%
  \BibitemOpen
  \bibfield  {author} {\bibinfo {author} {\bibfnamefont {Jaroslav}\ \bibnamefont {Adam}} \emph {et~al.} (\bibinfo {collaboration} {ALICE}),\ }\bibfield  {title} {\enquote {\bibinfo {title} {{Centrality dependence of inclusive J/\ensuremath{\psi} production in p--Pb collisions at $ \sqrt{s_{\mathrm{NN}}}=5.02 $ TeV}},}\ }\href {\doibase 10.1007/JHEP11(2015)127} {\bibfield  {journal} {\bibinfo  {journal} {JHEP}\ }\textbf {\bibinfo {volume} {11}},\ \bibinfo {pages} {127} (\bibinfo {year} {2015})},\ \Eprint {http://arxiv.org/abs/1506.08808} {arXiv:1506.08808 [nucl-ex]} \BibitemShut {NoStop}%
\end{thebibliography}%

\end{document}